\documentclass[12pt,preprint2]{aastex}

\usepackage{emulateapj5}
\usepackage{apjfonts}
\usepackage{onecolfloat5}

\newcommand{\tskip}{\omit\tablevspace{1pt}}
\renewcommand\colhead[1]{\normalsize{#1}}

\renewcommand\plotone[1]{\epsffile{#1}}
\renewenvironment{figure}{\epsfxsize 3.5in}{\vskip 12pt}

\shorttitle{2MASS QSO Survey}
\shortauthors{Marble et al.}

\submitted{To appear in {\em The Astrophysical Journal}}

\begin{document}
\twocolumn[
\title{An HST/WFPC2 Snapshot Survey of 2MASS-Selected Red QSOs}

\author{Andrew R. Marble, Dean C. Hines, Gary D. Schmidt \& Paul S. Smith}
\affil{Steward Observatory, University of Arizona, Tucson, AZ 85721}
\author{Jason A. Surace \& Lee Armus}
\affil{SIRTF Science Center, California Institute of Technology, Pasadena, CA  91125}
\author{Roc M. Cutri \& Brant O. Nelson}
\affil{IPAC, California Institute of Technology, Pasadena, CA 91125}

\vskip 12pt

\begin{abstract}

Using simple infrared color selection, the Two Micron All Sky Survey (2MASS)
has found a large number of red, previously unidentified, radio-quiet
quasi-stellar objects (QSOs).  Although missed by UV/optical surveys, the
2MASS QSOs have $K_S$-band luminosities that are comparable to
``classical'' QSOs.  This suggests the possible discovery of a previously predicted
large population of dust-obscured radio-quiet QSOs.  We present the results
of an imaging survey of 29 2MASS QSOs observed with the
Wide-Field Planetary Camera 2 onboard the \emph{Hubble Space Telescope}.
$I$-band images, which benefit from the relative faintness of the nuclei at 
optical wavelengths, are used to characterize the host galaxies, measure the 
nuclear contribution to the total observed $I$-band emission, and to survey
the surrounding environments.  The 2MASS QSOs are found to lie
in galaxies with a variety of morphologies, luminosities, and dynamical 
states, not unlike those hosting radio-quiet PG QSOs.
Our analysis suggests that the extraordinary red colors of the
2MASS QSOs are caused by extinction of an otherwise typical QSO spectrum 
due to dust near the nucleus.

\end{abstract}

\keywords{quasars: general --- galaxies: active --- galaxies: interactions --- dust, extinction}

]

\section{Introduction}\label{sec_intro}

Mostly immune from absorption processes and obscuration by dust, radio surveys are 
ideal for identifying complete samples of quasi-stellar objects (QSOs).
However, only $\sim$10\% of QSOs are ``radio-loud''\footnote{Radio-loud QSOs
are distinguished by their radio to optical flux ratio, $R \ge 10$. This 
cut-off value, as well as the wavelengths used, vary throughout the 
literature.}, necessitating the use of other more vulnerable wavelengths.
Optical surveys designed to find UV-excess
objects have long prevailed.  Despite their success, searches in the 
optical are fated to miss QSOs that might be reddened by dust.

\citet{web95} studied a large sample of radio-selected quasars and found
an enormous amount of scatter in their $B_J-K$ colors ($1<B_J-K<8$).  This was interpreted
as the result of varying amounts of dust within the quasar host galaxy.  Optically-selected
QSOs showed a much smaller color range, presumably because the redder objects
were not detected.  If the dust content around radio-loud and radio-quiet
QSOs does not differ significantly, their results suggested that approximately 
80\% of radio-quiet QSOs are missed by optical surveys.  

The prediction of a large, predominantly undetected population of red QSOs is 
consistent with theoretical
explanations for the hard x-ray background \citep{com95}.  However, some
have claimed that the large spread in $B_J-K$ colors seen by
\citet{web95} is not due entirely to dust obscuration \citep{boy95,ben98}.  
In that case, any missing population of QSOs would be much smaller, if
even significant.

The discovery of highly reddened and buried QSOs in the \emph{Infrared Astronomical
Satellite} (\emph{IRAS}) database
\citep[e.g.,][]{bei86,low88,cut94,hin95,hin99a,hin99b}
confirmed that at least some radio-quiet QSOs are
indeed obscured along our line of sight by dust.
While \emph{IRAS} was only able to detect the mid and far-infrared emission of
extremely reddened QSOs, it was quickly recognized that
a wide area, near-infrared survey could potentially uncover
large numbers of previously unknown QSOs due to less extinction at
2$\micron$ than in the optical.

With this key feature in mind, \citet{cut01} are compiling a
catalog of active galactic nuclei (AGN)
derived from the Two
Micron All-Sky Survey \citep[2MASS:][]{skr97} Point Source Catalog.  
Even with a relatively small fraction of the sky analysed
so far, spectroscopic followup has confirmed over 400
previously unidentified emission-line AGN with QSO-like luminosities
\citep[$M_{K_{s}} \le -23$ and $z \lesssim 0.4$:][]{cut01,cut02}.

The nature of these red 2MASS QSOs is an open question.
The $K\/$-band and [O~III]$\lambda$5007 luminosities of many of the newly
found objects fall in the same ranges as for UV/optically-selected QSOs \citep{smi03}.
However, nearly all have much redder colors ($B-K_S$) compared with
traditional QSOs \citep[e.g., PG QSOs:]
[]{sch83}, implying that many of the 2MASS objects are obscured
from our direct view by dust.  \citet{smi02} find that at least 10\% of the 2MASS
QSOs are ``highly-polarized'' ($P>3\%$); a significantly higher
fraction than both the PG \citep{ber90} and broad absorption line
QSOs \citep{sch99}.  Indeed, only the \emph{IRAS}-selected hyperluminous
infrared galaxies (HIGs) show higher polarization.  In addition, several of the highly
polarized 2MASS QSOs exhibit broad polarized emission lines even
though they are dominated by narrow emission lines in total flux
\citep{sch02,smi03}. This is very reminiscent of the simple
orientation-dependent unified schemes for Seyfert galaxies 
\citep{ant93,tra95} and HIGs
\citep{hin93,hin95,hin99a,hin99b,goo96,tra01}.
In these schemes, a dusty torus obscures our direct view but light
from the nucleus escapes through the open poles to be scattered (thus
polarized) into our line of sight; type 2 objects appear different
from type 1 simply because their tori are more highly inclined.

However, there is still an ambiguity between simple orientation
schemes and those that involve significant dust cover from debris left by
strong galaxy-galaxy interactions.
It has been suggested that ultraluminous infrared galaxies (ULIRGs),
\emph{IRAS}-selected systems with the bulk of their energies emerging in the
far-infrared and with bolometric luminosities of
$10^{12}-10^{13}$L$_{\odot}$, can evolve into classical QSOs once the gas
and dust around their nuclei are either consumed as fuel by a powerful
starburst or a nascent AGN, or blown away in powerful winds \citep{san88,hec90}.
In this scenario, some or all of the 2MASS QSOs might be aging ULIRGs, complete
with post-starburst stellar populations, fading tidal remnants,
disturbed morphologies, and perhaps closely spaced multiple nuclei.

The 2MASS-selected QSOs might therefore be (1)
relatively normal QSOs seen from near the plane of an obscuring torus,
(2) fairly young, evolving post-ULIRG systems wherein most of the
molecular gas and dust has been removed from the circumnuclear
environment, but that still retain enough material to make them
``invisible'' to classical UV or visual searches for AGN, or (3) an entirely
new aspect of the overall phenomenon of nuclear activity.  As part of a 
multi-wavelength investigation of the
2MASS AGN, we have performed a snapshot imaging survey with the Wide-Field Planetary
Camera 2 (WFPC2) aboard
the \emph{Hubble Space Telescope (HST)} to investigate these 
possibilities. The images have
enabled us to characterize the host galaxies, estimate the relative
contributions to the emitted optical flux from the host galaxy and the
active nucleus, and survey the immediate environments of these
objects.  

\section{Data}\label{sec_data}

\subsection{The 2MASS QSO Sample}

Candidate AGN were selected 
from 2MASS according to color ($J-K_{S}>2$), galactic latitude ($|b|>30\degr$), 
and detection (complete to $K_S\leq15.0$) in each of the three 2MASS bands, $JHK_{S}$ \citep{cut01,cut02}. These 
criteria efficiently filter almost all known AGN and all but the coolest stars \citep{bei98}. 
Approximately 80\% of these candidates were confirmed as AGN via follow-up
optical spectroscopy that also determined their redshifts and spectral type.
Of these AGN, most
have absolute $K_S$ magnitudes that easily fall within the luminosity 
range of UV/optically-selected QSOs (Figure~\ref{fig_pg}). A representative 
subset (90) of this 
red, radio-quiet QSO ($M_{K_S}<-23$) catalogue was surveyed for optical broad-band polarization 
\citep{smi02}.  Twenty-nine of these objects comprise the sample addressed by 
this paper. They range in redshift from $z=0.136$ to $z=0.596$ with a median 
value of 0.213.  The sample members and their previously determined properties 
are provided in Table~\ref{tbl_sample}.  
For brevity, the designation of objects 
from the 2MASS Incremental Release Point Source Catalog has been shortened to 
\emph{2Mhhmmss$\pm$ddmm} in the text.

\subsection{Observations}\label{sec_obs}

A WFPC2 snapshot survey was conducted between the UT dates of 2 July 2001
and 14 October 2001 during Cycle 10 of the \emph{HST}
observing schedule. Twenty-eight of fifty-four potential targets were
observed during that period.  A final twenty-ninth object (2M010607+2603) was added 
on 10 June 2002.  The QSOs were centered on the planetary camera (PC) and observed
with a single filter, F814W. Each observation
consisted of two 400 second integrations in CR-SPLIT mode to allow for the
removal of cosmic rays.  In order to maximize integration time, no dithering 
was attempted.  The survey was designed to balance the scientific
requirements of the project with the simplicity needed for successful
snapshot scheduling. WFPC2 was selected to facilitate comparisons
with previous studies, and the PC, with 0.046$\arcsec$
pixels and a 36.8$\arcsec$ field of view, was chosen over the
wide field cameras in order to better sample the point spread function (PSF).
As a result, the light contribution of AGN can be subtracted more
precisely, but at the expense of sensitivity to extremely low surface
brightness features. The F814W filter, which corresponds closely to the
Cousins $I$-band (hereafter referred to simply as I), has the dual advantage of 
being sensitive to red light from older stellar populations, while still 
benefitting from the relative faintness of the nuclei at optical wavelengths.

\vskip 12pt
\begin{figure}
\centering
\plotone{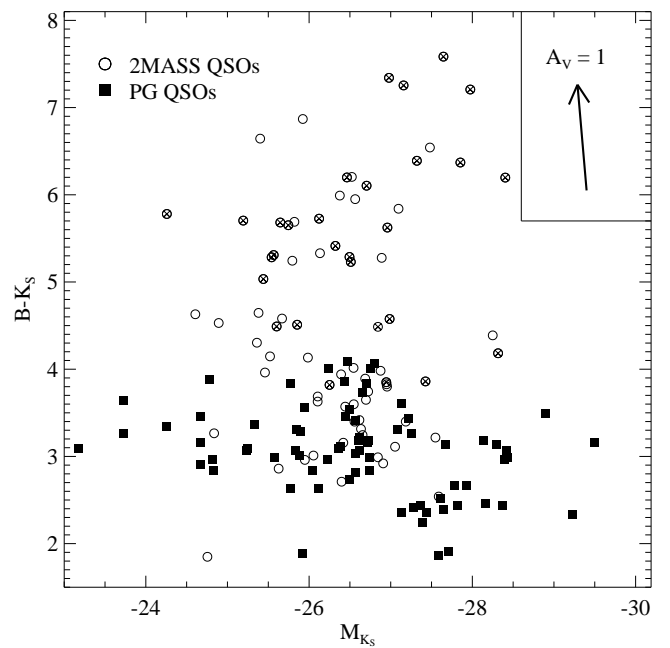}
\vskip 6pt
\figcaption{B-K$_S$ color vs. absolute K$_S$ magnitude.  The 2MASS QSOs are much redder
than the UV-excess PG QSOs but have comparable near-infrared luminosities \citep{neu87}. 
The circles are a representative inventory of the red QSOs 
discovered by 2MASS.  Those investigated in this paper are indicated by $\otimes$.  
The PG QSOs plotted here are the 75 (out of 114) with $M_K<-23$ and $z<0.6$.
The reddening vector shown in the upper righthand corner represents one magnitude of 
extinction in the $V$-band \citep[$A_B=1.324$, $A_K=0.112$:][]{rie85}.}
\end{figure}

\begin{table*}[t]
\begin{center}
\vskip 12pt
{\sc Table \ref{tbl_sample}\\
\smallskip\hbox to\hsize{\hfil{Sample of 29 2MASS-selected red QSOs}\hfil}}
\small
\setlength{\tabcolsep}{6pt}
\begin{tabular}{ccccccccc}
\tskip \tableline \tskip
\colhead{Object} & \colhead{RA} & \colhead{Dec} & 
\colhead{$z$\tablenotemark{a}} & \colhead{$K_S$\tablenotemark{b}} & 
\colhead{$J-K_S$\tablenotemark{b}} & \colhead{$J-H$\tablenotemark{b}} & 
\colhead{$M_{K_S}$\tablenotemark{c}} & \colhead{Type\tablenotemark{d}} \\
\colhead{(2MASSI J)} & \colhead{(J2000)} & \colhead{(J2000)} & 
\colhead{} & \colhead{} & \colhead{} & \colhead{} & \colhead{} & \colhead{} \\
\tskip \tableline
\tableline \tskip
000810.8+135452 & 00 08 12.26 & 13 55 14.68 & 0.185 & 14.4 &  2.1 &  0.8 & -25.2 & 2 \\
005055.7+293328 & 00 50 57.15 & 29 33 51.82 & 0.136 & 13.2 &  2.1 &  1.0 & -25.6 & 2 \\
010607.7+260334 & 01 06 09.89 & 26 03 43.07 & 0.411 & 14.6 &  2.7 &  1.2 & -27.3 & 1.2 \\
015721.0+171248 & 01 57 22.66 & 17 13 08.17 & 0.213 & 13.2 &  2.7 &  1.3 & -27.0 & 1.x \\
022150.6+132741 & 02 21 52.64 & 13 27 46.94 & 0.140 & 13.2 &  2.4 &  1.1 & -25.7 & 1.8 \\
023430.6+243835 & 02 34 32.33 & 24 38 55.20 & 0.310 & 13.7 &  2.2 &  1.1 & -27.2 & 1.5 \\
032458.2+174849 & 03 25 00.27 & 17 48 59.48 & 0.328 & 12.8 &  2.4 &  1.0 & -28.3 & 1 \\
034857.6+125547 & 03 48 59.61 & 12 55 56.82 & 0.210 & 13.6 &  3.3 &  1.5 & -26.7 & 1.x \\
092049.0+190320 & 09 20 51.16 & 19 03 12.47 & 0.156 & 14.9 &  2.1 &  1.2 & -24.3 & 1.b \\
125807.4+232921 & 12 58 05.30 & 23 29 27.66 & 0.259 & 13.4 &  2.1 &  0.9 & -26.9 & 1 \\
130700.6+233805 & 13 06 58.46 & 23 38 02.45 & 0.275 & 13.4 &  3.3 &  1.6 & -27.6 & 1.b \\
145331.5+135358 & 14 53 29.47 & 13 53 52.13 & 0.139 & 13.1 &  2.2 &  1.0 & -25.9 & 1.x \\
151621.1+225944 & 15 16 19.13 & 22 59 31.25 & 0.190 & 14.1 &  2.1 &  1.0 & -25.5 & 1.b \\
152151.0+225120 & 15 21 49.08 & 22 51 07.62 & 0.287 & 14.3 &  2.3 &  0.9 & -26.5 & 1.x \\
154307.7+193751 & 15 43 05.68 & 19 37 44.90 & 0.228 & 12.7 &  2.3 &  1.1 & -27.4 & 1.5 \\
163700.2+222114 & 16 36 58.71 & 22 21 36.03 & 0.211 & 13.6 &  2.1 &  1.1 & -26.3 & 1.5 \\
163736.5+254302 & 16 37 34.36 & 25 42 54.13 & 0.277 & 14.2 &  2.3 &  1.2 & -26.5 & 1.9 \\
165939.7+183436 & 16 59 37.68 & 18 34 30.53 & 0.170 & 12.9 &  2.2 &  0.9 & -26.5 & 1.5 \\
170003.0+211823 & 17 00 01.09 & 21 18 37.62 & 0.596 & 14.9 &  2.5 &  1.4 & -28.0 & 1.5 \\
171442.7+260248 & 17 14 40.52 & 26 02  49.5 & 0.163 & 13.1 &  2.2 &  1.1 & -26.3 & 1 \\
171559.7+280717 & 17 15 58.64 & 28 07 43.72 & 0.524 & 14.6 &  2.5 &  1.3 & -27.9 & 1.8 \\
222202.2+195231 & 22 22 03.14 & 19 52 58.93 & 0.366 & 13.3 &  2.9 &  1.3 & -28.4 & 1.5 \\
222221.1+195947 & 22 22 20.86 & 20 00 17.54 & 0.211 & 12.9 &  2.1 &  0.9 & -27.0 & 1.5 \\
222554.2+195837 & 22 25 55.00 & 19 59 05.57 & 0.147 & 13.5 &  2.1 &  1.0 & -25.6 & 2 \\
225902.5+124646 & 22 59 03.97 & 12 47 09.33 & 0.199 & 14.1 &  1.9\tablenotemark{e} &  0.9 & -25.6 & 1.x \\
230304.3+162440 & 23 03 05.71 & 16 25 03.07 & 0.289 & 14.7 &  2.3 &  1.4 & -26.1 & 2/S \\
230442.4+270616 & 23 04 43.80 & 27 06 40.73 & 0.237 & 14.8 &  2.1 &  1.3 & -25.4 & 1.5 \\
232745.6+162434 & 23 27 46.26 & 16 25 02.68 & 0.364 & 14.5 &  2.4 &  1.1 & -27.0 & ? \\
234449.5+122143 & 23 44 51.18 & 12 22 02.48 & 0.199 & 12.9 &  2.1 &  0.9 & -26.8 & 1 \\
\tskip\tableline
\end{tabular}
\vspace*{-12pt}
\tablenotetext{a}{Redshifts from \citet{cut01}.}
\tablenotetext{b}{Near-infrared apparent magnitudes and colors from the 2MASS Point Source Catalog.}
\tablenotetext{c}{Calculated (Eq.~\ref{eq_abs}) from Column 5 values ($H_0=75$ km s$^{-1}$ Mpc$^{-1}$ and $q_0=0.5$).}
\tablenotetext{d}{Spectral classifications from \citet{cut01} and \citet{smi03}.}
\tablenotetext{e}{Following recent observations, this object no longer meets the J-K$_S>2$ selection criterion.  It \emph{has} been included in the analysis of the sample.}
\end{center}
\end{table*}

\subsection{Data Reduction}\label{sec_red}

The WFPC2 images were calibrated using the STSDAS routine \emph{calwp2}
that is incorporated into the STScI data reduction pipeline and makes use
of the best reference files available. Subsequently, a
flat-field correction was applied as described in the WFPC2 Instrument
Science Report 2001-07 \citep{kar01}.  The two images
for each object were combined and cosmic rays were removed using the IDL routine
 \emph{cr\_reject} that is nearly identical to \emph{crrej} in IRAF and
is available in the IDL Astronomy User's Library.  No compensation was made for
charge transfer efficiency (CTE) problems known to affect WFPC2, resulting in an
underestimation of the illumination in the center of the chip at the few percent
level \citep{whi97,bir02}.  The dominant source of error in the magnitudes we 
derive is not the CTE compensation, but uncertainty in the PSF-subtraction.  Sky 
subtraction was carried out using custom IDL software.

\subsection{PSF-Subtraction}\label{sec_psf}

Factors affecting the PSF include the 
filter response, scattering due to surface irregularities in the optics such 
as dust, the shape of an object's spectral energy distribution (SED), and especially
slight changes in the position of the secondary mirror, or ``breathing''.
While all of these 
may vary slightly between observations, their combined effect is minimized by 
creating a PSF directly from observed point sources.
Using the sample QSOs themselves for this 
purpose is preferable to field stars for two reasons. The PSF is 
somewhat dependant on a source's position on the chip, and a 
star's color significantly affects whether or not it is suitable for 
modeling a QSO PSF \citep{bah97}.  As this observing program was 
designed to be a single filter snaphot survey, colors were not available and valuable 
orbits were not dedicated to observing stars centrally located on the 
chip. 

In order to identify the most ``point source like'' QSOs in our sample,
an artificial PSF created 
by the software package \emph{Tiny TIM} \citep{kri01} was scaled and 
subtracted from each image where a host galaxy was not 
readily apparent. 
Direct comparison of the resulting profiles revealed extended emission 
around six of the nine candidates.  Two of the remaining three were 
not optimal for the purpose of constructing a PSF. 2M170003+2118 was 
relatively faint requiring that it and its inherent noise be scaled 
significantly to combine it with brighter objects. A saturated star 
in the field of 2M023430+2438 was a source of contamination that 
could not be removed reliably. Therefore, our PSF was created solely 
from 2M222202+1952. 

The light contribution of the AGN can be bounded by considering two extreme 
cases.  The upper limit to the AGN emission corresponds to the PSF amplitude 
which results in the subtraction residuals summing to zero within a specified aperture 
(defined here as the full width at half maximum (FWHM) of the PSF).
With the constraint that the PSF amplitude must be sufficiently 
large to eliminate the presence of diffraction spikes and an Airy pattern,
the lower limit corresponds to the removal of the central peak 
without introducing any negative residuals.

Between these bounds, the PSF amplitude which minimizes the variance 
of the residuals can be considered the best determination of the light
contribution of the AGN.  This approach yields generous errors.  Finally, for 
the purpose of estimating a $3\sigma$ flux uncertainty due to 
PSF-subtraction, the smaller of the differences between the accepted value 
and the upper and lower limits was adopted.

\vskip 24pt

\subsection{QDOT ULIRGs}\label{sec_ulirgs}

Before presenting the results from the analysis of the 2MASS QSOs, we 
briefly describe a second sample of objects investigated in this paper.
For reasons outlined in \S \ref{sec_intro} we wished to
compare properties of the 2MASS QSOs to a sample of ULIRGs.  Identification
of companion objects surrounding the latter are not currently available 
in the literature, requiring us to perform our own analysis of \emph{HST} 
archival data.  A substantial number of ULIRGs have been observed with 
WFPC2/F814W \citep{bor00} and are publicly available in 
the archive.
These images were taken in exactly the same manner as the 2MASS QSO 
sample, with the exception of being centered on WF3
instead of the PC.  Cross-listing those objects with the 
QDOT\footnote{QDOT is a redshift survey of 2,387 \emph{IRAS} galaxies
brighter than the \emph{IRAS} PSC $60\micron$ completeness limit \citep[and references therein]{law99}.  
In the context of this paper,
it is only used as means for constructing a ULIRG sample with uniform
selection criteria.}
ULIRG sample 
and imposing the same sky position restriction used to select the 2MASS 
QSOs ($|b|>30^{\circ}$) results in a homogeneous sample of 30 high-latitude 
ULIRGs.  These objects and their redshifts are  
presented in \S \ref{sec_env}.  Precisely the same steps described 
in \S \ref{sec_red} were applied to these observations.

\vskip 24pt

\section{Results}\label{sec_results}

Images of all 29 2MASS QSOs prior to and following PSF-subtraction are 
shown in Figure~\ref{fig_hosts1}.  The entire sample is displayed at the same
angular and physical scales in the first and second columns, respectively.  
The second and third columns show the objects before and after 
subtraction of the light contribution from the AGN.  In some cases this
difference is dramatic; in others there is very little visual change.
The position angle of polarization (taken from \citet{smi02,smi03}) for those objects
with measured $P/\sigma_P\ge4$ is denoted by the double-headed arrow in the upper 
righthand corner in column two. Isophotes of the residual 
images (and the masks used in calculating them) are shown in the last column.
In some cases, the isophotes more clearly reveal underlying structure,
highlighting
features such as multiple nuclei and isophotal twisting
caused by interaction or the presence of spiral arms.  The solid 
contours indicate flux levels that are distinguishable from image noise
at the $3\sigma$ confidence level, as opposed to the dotted contours.  
The dashed line identifies the 
isophote used to determine the eccentricity and position angle of the 
host galaxy.   Measured properties of the host galaxies
are listed in Table~\ref{tbl_properties} and are discussed below.
Additional notes on individual objects can be found in Appendix~\ref{app_notes}.

\vskip 36pt

\subsection{Luminosities}\label{sec_lum}

The PSF-subtraction revealed the relative flux contributions of the AGN and host 
galaxies within $10\arcsec$ radius apertures\footnote{Foreign sources
such as stars, background galaxies and companions were masked.}.  
Absolute magnitudes were then calculated using the standard formula \citep{pee93}
\begin{equation}\label{eq_abs}
m - M = 42.38 + 5 \log[z/h] + 1.086(1-q_0)z + K
\end{equation}
and the prescription found in the WFPC2 Instrument Handbook \citep{bir02}
\begin{equation}\label{eq_app_I}
m_I=-2.5 \log\left[2.508\times10^{-18} \left(DN / Exp\right)\right]-22.31
\end{equation}
where $h=H_0/100$, $K$ is the redshift-dependant K-correction, $DN$
is the number of counts in the F814W filter, and $Exp$ is the exposure time.
Throughout this paper, values for the cosmological constants $q_0$ and $H_0$ 
are taken to be 0.5 and 75 km s$^{-1}$ Mpc$^{-1}$, respectively.  Although 
transformation between the \emph{HST} system and standard magnitudes generally 
depends strongly on the shape of an object's spectrum, the effect is minimal 
for the F814W filter ($\sim$0.05 magnitudes). Equation~\ref{eq_app_I} yields 
results nearly identical to those obtained by \citet{hol95}.

K-corrections for the host galaxies, ranging between $-0.21$ and $-1.01$ magnitudes,
were measured from spiral and elliptical SED templates.  For those 
hosts with ambiguous morphologies, an average of the two templates was
used.  For the AGN, K-corrections were taken from the Sloan Digital Sky Survey 
(SDSS) composite QSO spectrum \citep{van01}.  Of course, the 2MASS QSOs are
much redder than UV/optically selected QSOs and their physical nature has yet
to be adequately determined.  Extrapolating their SED from a ``typical'' 
QSO spectrum in order to calculate K-corrections is not necessarily 
correct.  Fortunately, for the redshift range of this sample, these corrections are 
relatively small ($0.14<K<0.28, \langle K \rangle =0.18$).  

Absolute magnitudes for the host galaxies and AGN are given in 
Table~\ref{tbl_properties}.  For the latter, both absolute rest frame $I$-band and 
apparent
(without K-corrections) F814W values are listed.  In this cosmology, an L* galaxy
has an $I$-band absolute magnitude of $M_I\sim-22.3$ \citep{nor02,bla03}.
The 2MASS hosts range between 
L$\sim0.3$L* and L$\sim4$L* with three quarters of them being brighter than L*.
The range in luminosity of the nuclei extends over 7 magnitudes, with only ten of the 29
objects satisfying the traditional $M_B<-21.5+5\log h$ 
criterion\footnote{Extrapolating to $I$-band yields $M_I<-22.3$.} for QSO 
classification \citep{sch83}.  Section~\ref{sec_discussion} presents evidence that AGN 
luminosity is correlated with the $I_{NUC}-K_S$ color index, suggestive of a reddening effect.

\subsection{Surface Brightness Profiles}\label{sec_profiles}

Isophotes were fit to the host galaxy images after the point source was 
subtracted and nearby stars/galaxies were masked.  The methodology used is 
described by \citet{buo94} and references therein.
Surface brightness profiles were then constructed using these
isophotes. The average magnitude per square arc second was calculated
as a function of radius along the semi-major axis for the regions between
successive elliptical annuli. Figure~\ref{fig_profiles1} shows surface brightness
profiles for each object, plotted against $r$
as well as $r^{1/4}$. Linear profiles on the left reflect
an exponentially decreasing surface brightness (a disk). On the
right, linear profiles indicate a de Vaucouleurs $r^{1/4}$ law
that is characteristic of spiral bulges and elliptical galaxies.

\begin{figure*}
\centering
\includegraphics[scale=1.0,angle=0]{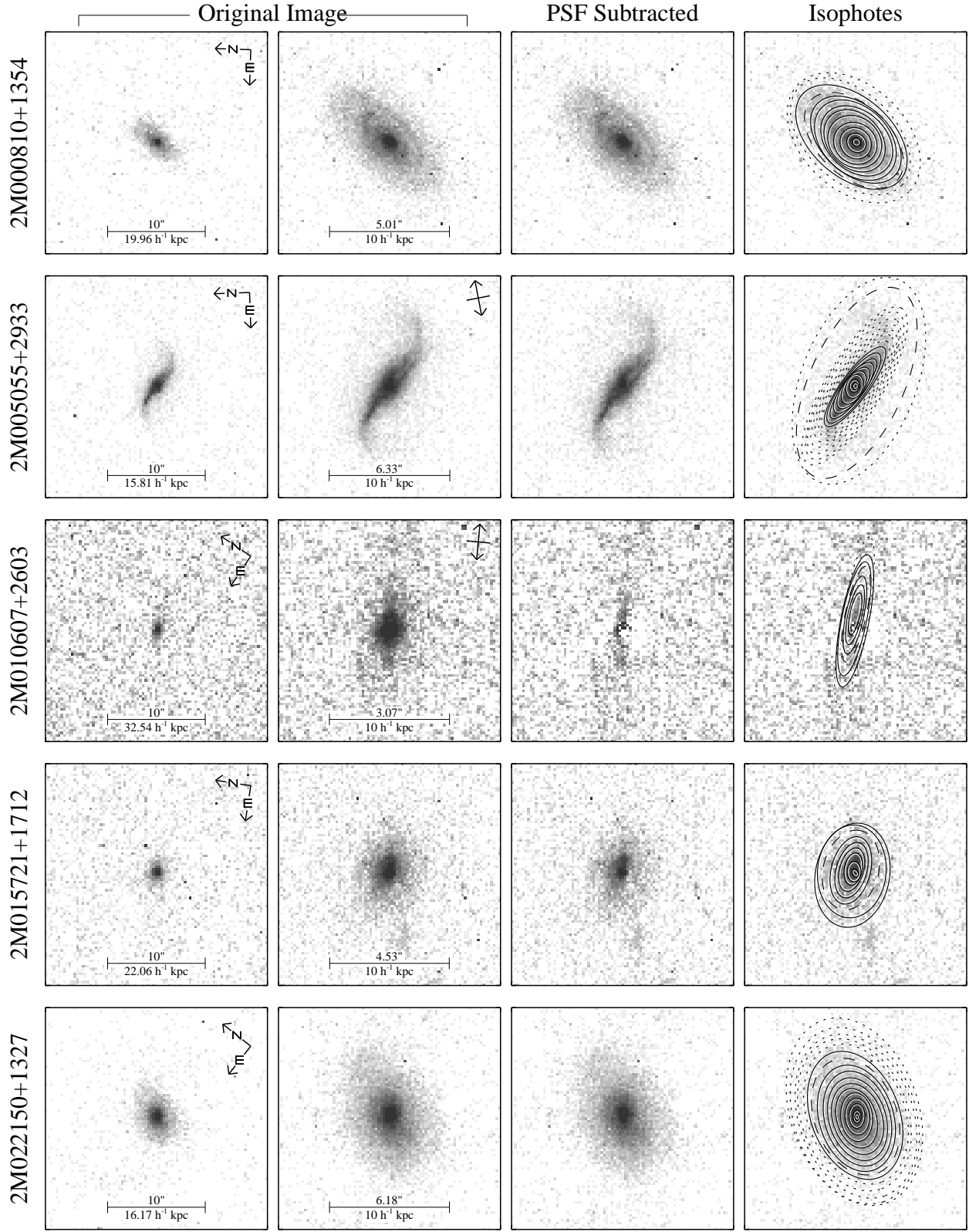}
\vskip 6pt
\caption{WFPC2 F814W images of 2MASS QSOs before (first and second columns) and after 
(third and fourth columns) PSF-subtraction.  The position angle of polarization for 
those objects with measured $P/\sigma_P\ge4$ is denoted by the double-headed arrow in 
the upper righthand corner in column two. Isophotes of the residual images (and the 
masks used in calculating them) are shown in the last column. The solid contours 
indicate flux levels that are distinguishable from image noise at the $3\sigma$ 
confidence level, as opposed to the dotted contours. The dashed line identifies the 
isophote used to determine the eccentricity and position angle of the host galaxy.}.
\end{figure*}
\begin{figure*}
\setcounter{figure}{1}
\centering
\includegraphics[scale=1.0,angle=0]{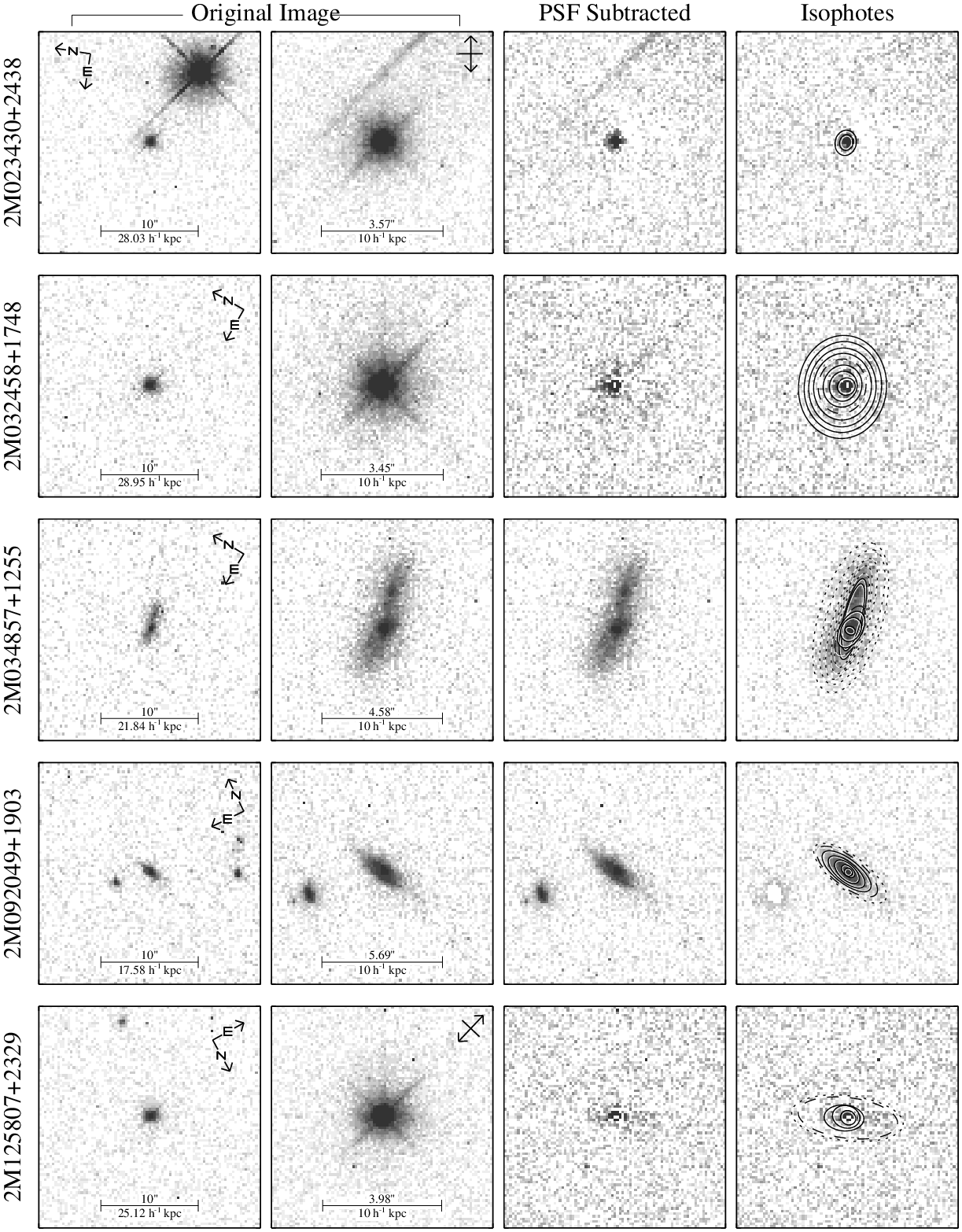}
\caption{\emph{continued...}}
\end{figure*}
\begin{figure*}
\clearpage
\setcounter{figure}{1}
\centering
\includegraphics[scale=1.0,angle=0]{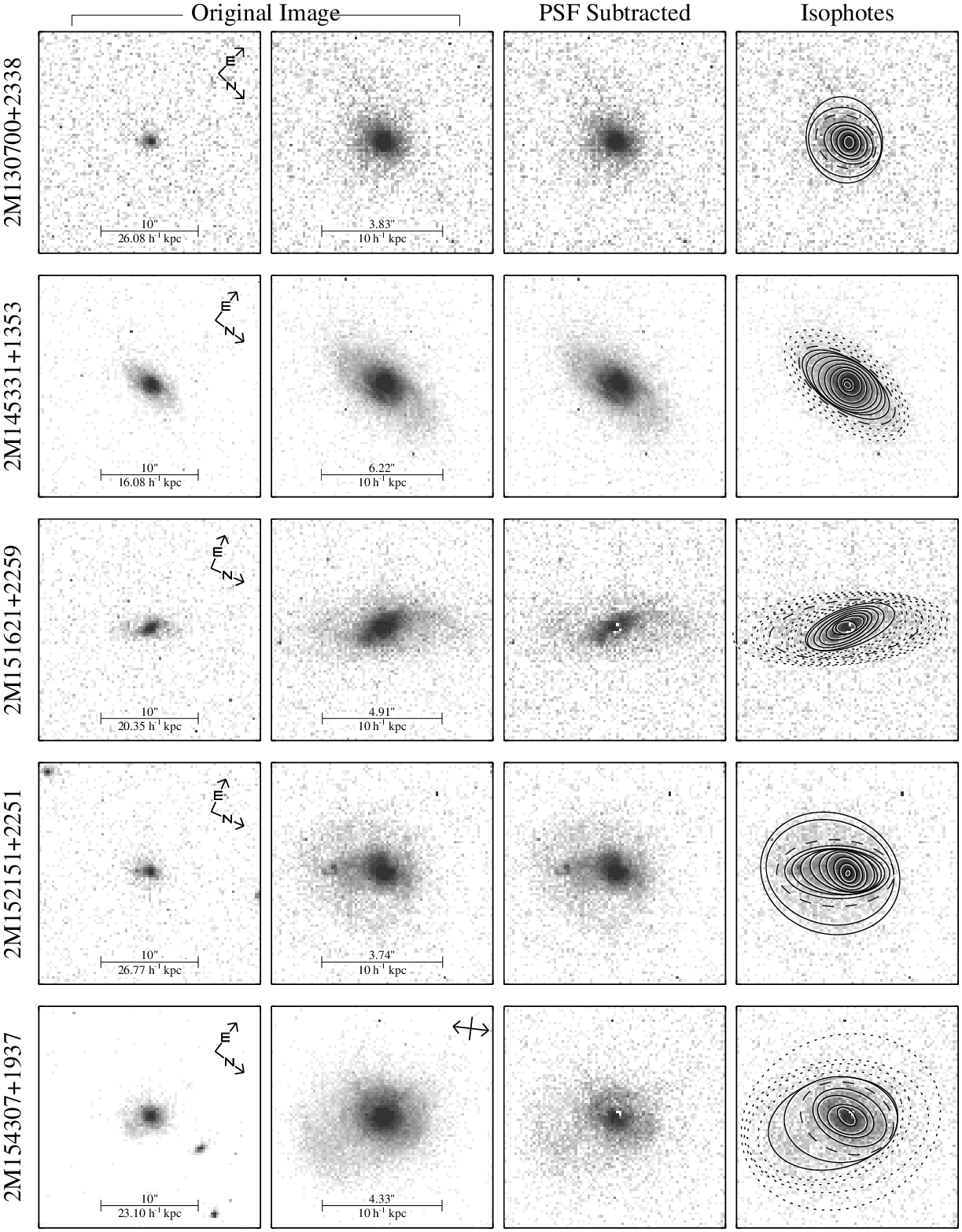}
\caption{\emph{continued...}}
\end{figure*}
\begin{figure*}
\setcounter{figure}{1}
\centering
\includegraphics[scale=1.0,angle=0]{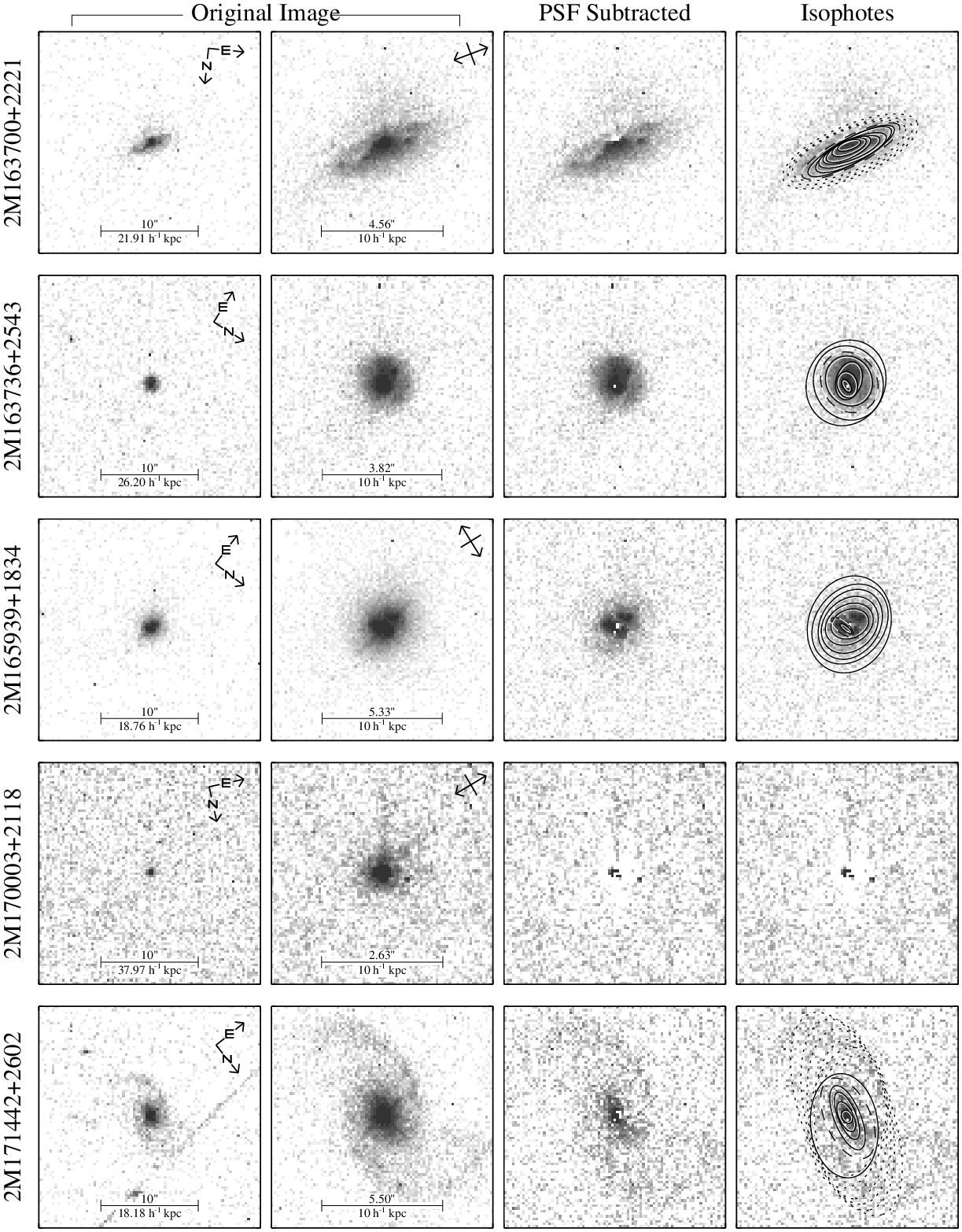}
\caption{\emph{continued...}}
\end{figure*}
\begin{figure*}
\setcounter{figure}{1}
\centering
\includegraphics[scale=1.0,angle=0]{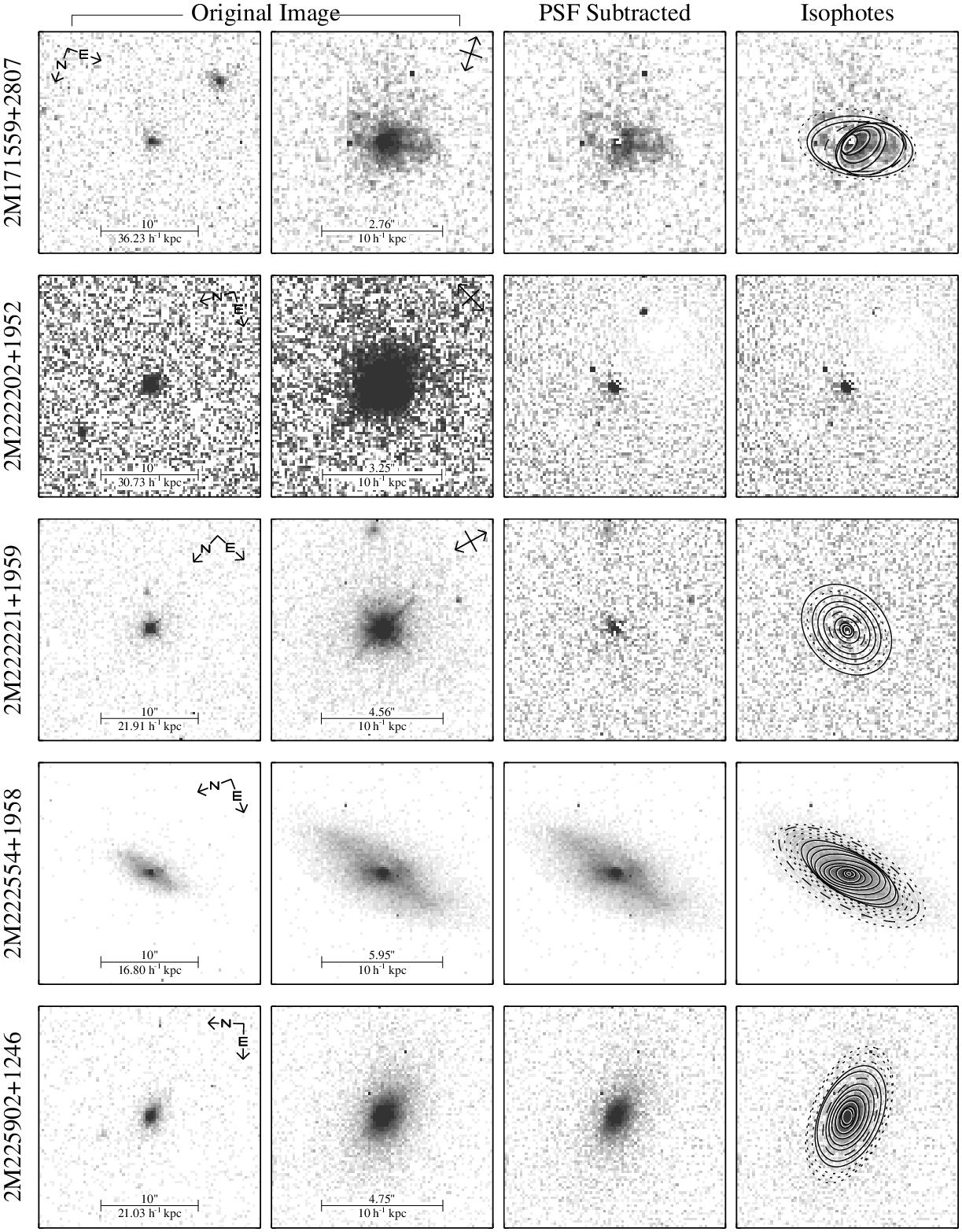}
\caption{\emph{continued...}}
\end{figure*}
\begin{figure*}
\setcounter{figure}{1}
\centering
\includegraphics[scale=1.0,angle=0]{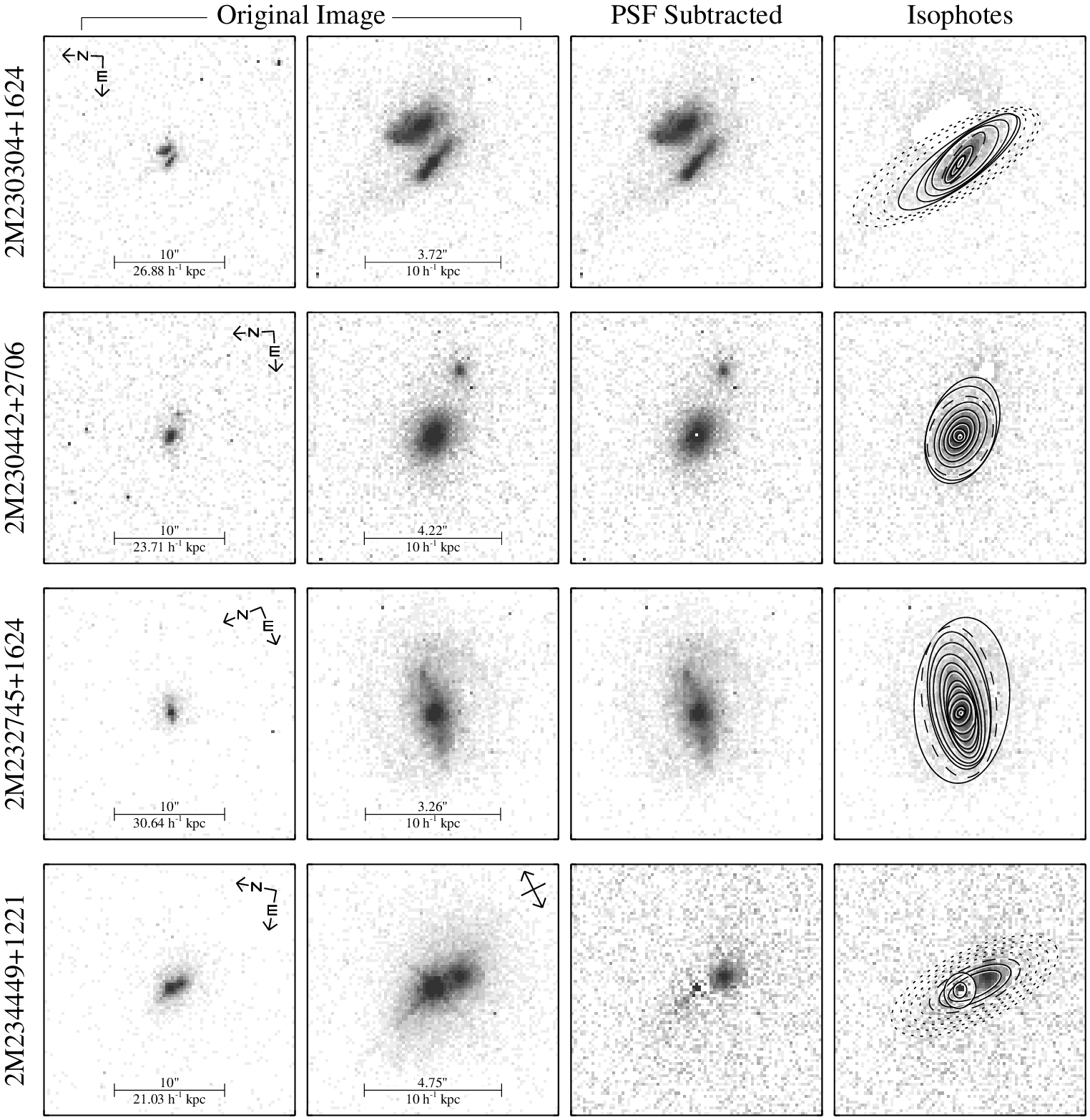}
\caption{\emph{continued...}}
\end{figure*}

\begin{figure*} 
\centering
\includegraphics[scale=0.9,angle=0]{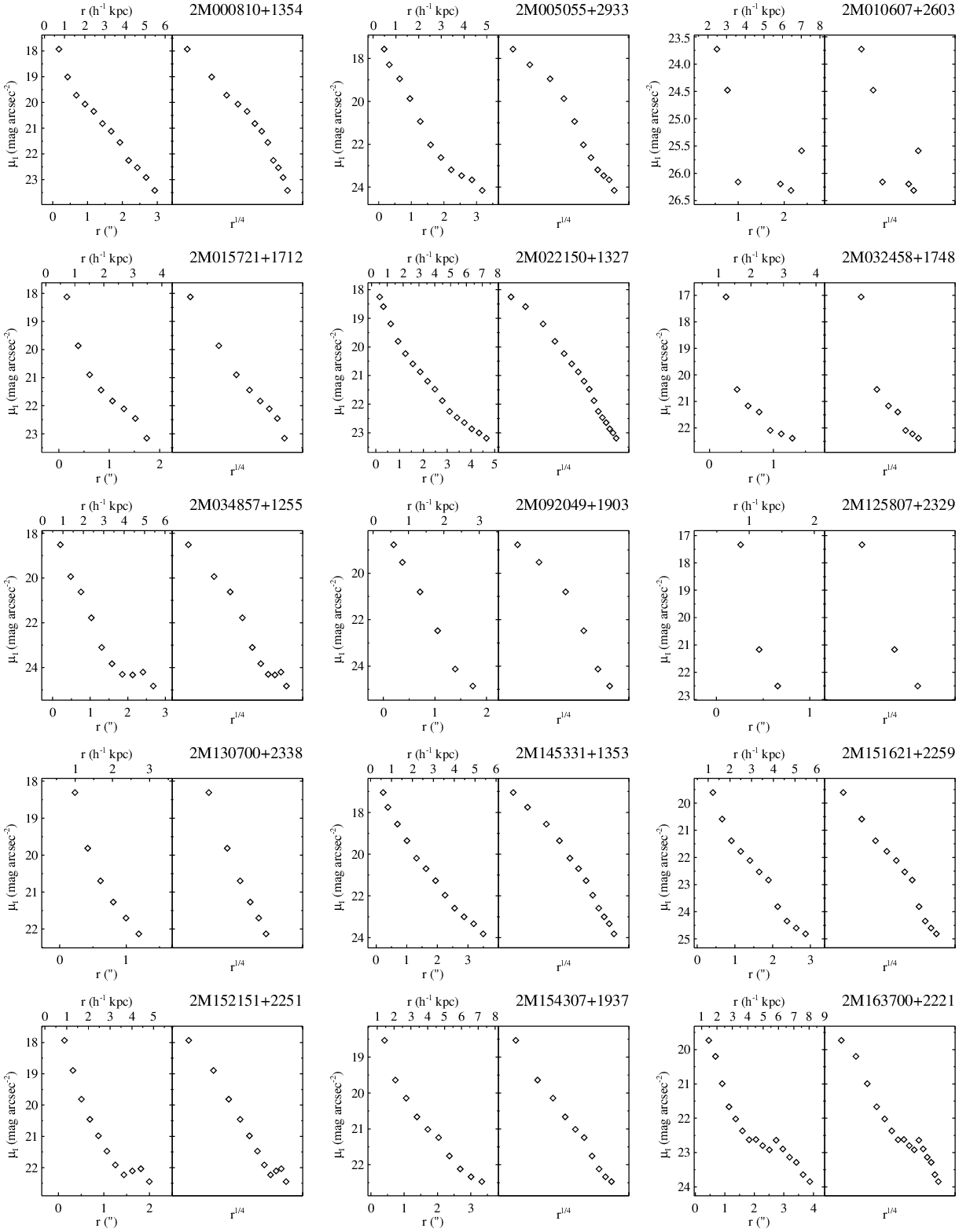}
\vskip 24pt
\caption{2MASS QSO host galaxy surface brightness profiles corresponding to the PSF-subtracted images shown in column three of Figure~\ref{fig_hosts1}.  A detailed account of these profiles is given in \S \ref{sec_profiles}.}
\end{figure*}
\begin{figure*}
\setcounter{figure}{2}
\centering
\includegraphics[scale=0.9,angle=0]{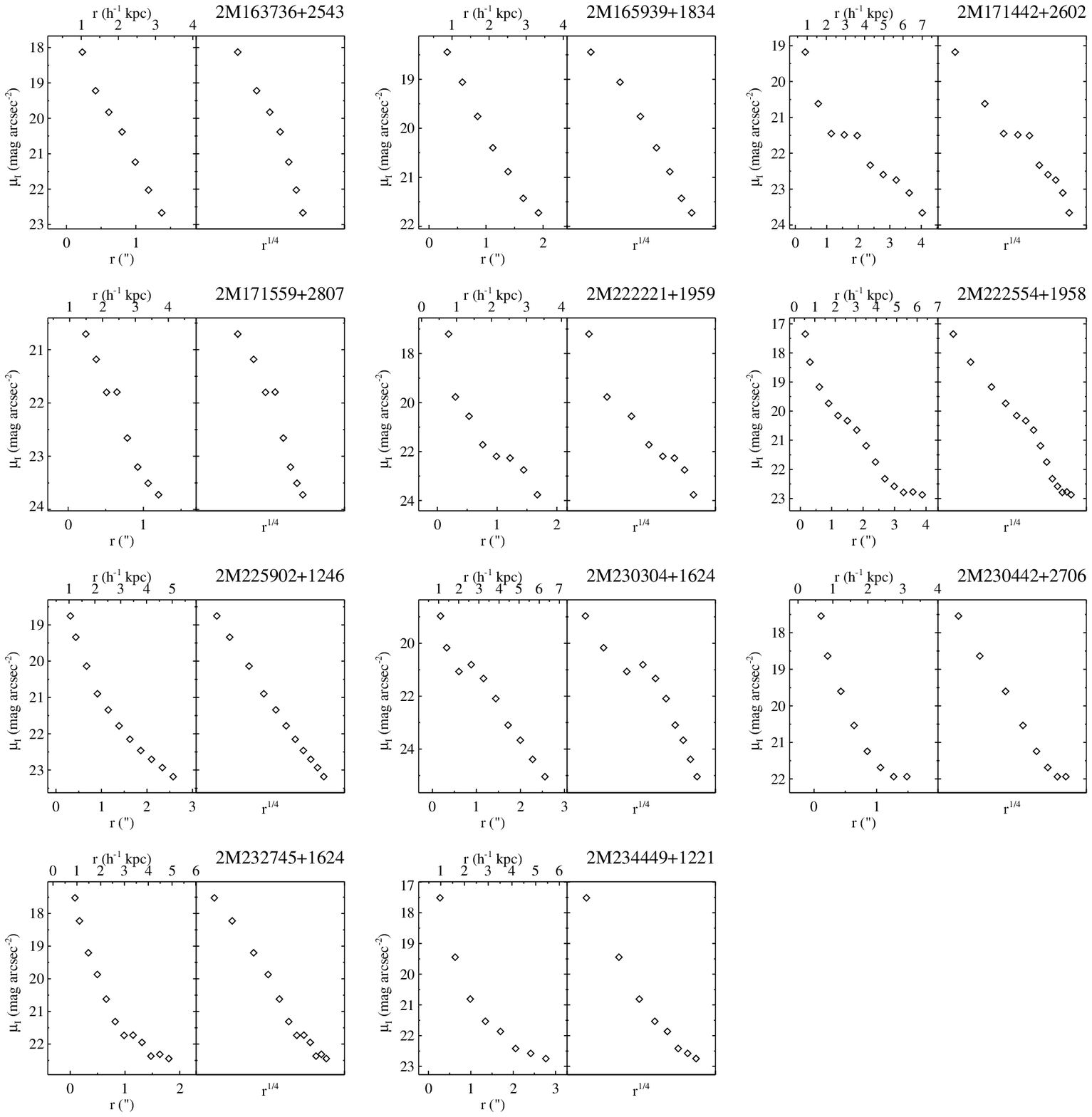}
\vskip 24pt
\caption{\emph{continued...}}
\end{figure*}

\noindent The profiles of the three objects with no obviously discernible host 
galaxies were noise-limited and are not included.  Additionally, it should be
noted that the profile shown for the last object, 2M234449+1221, may be
misleading.  The inner points are dominated by the immediate region 
surrounding the QSO, while at larger distances the surface brightness 
reflects the other feature clearly seen in Figure~\ref{fig_hosts1}.  
Whether this is a double nucleus within
the same host galaxy or a separate companion is unclear.  By chance this 
results in a fairly smooth surface brightness profile.

\subsection{Morphologies}

In three cases, a host galaxy 
could not be \emph{visually} distinguished from either the image noise or the 
PSF-subtraction residuals. Considering the survey's short exposure times, 
the PC's relative insensitivity to low surface brightness features, and the 
comparatively high redshifts ($z=0.310$, 0.366 and  0.596) of these 
objects, it is reasonable to assume that host galaxies are present but 
very faint\footnote{The 800 second observations reached a limiting magnitude 
of approximately $M_I<24$.}.  In fact, luminosities from the residual flux 
are consistent with the other host galaxies' absolute magnitudes. For the 
remainder of the sample, morphologies were assigned by visual 
inspection of both the PSF-subtracted images and the surface brightness 
profiles.  In many cases, classifying the type of host galaxy is an
extremely uncertain exercise.
Table~\ref{tbl_properties} lists only fairly unambiguous morphologies, leaving 
nearly half the sample unclassified.  Of the 26 QSOs with
visible hosts, nine reside in disk (spiral) galaxies and four are
found in what are well described as ellipticals.

The hosts classifed as spirals preferentially lie at low redshifts 
(Figure~\ref{fig_z}).  A Kolmogorov-Smirnov (KS) test yields only a $0.31\%$ 
probability that the redshifts of the spirals and remaining galaxies are 
drawn from the same parent population.  This suggests that some, or all, of 
the unclassified host galaxies may have spiral disks that are less obvious
due to their smaller angular extent and the shallow exposures of the shapshot 
survey.  This is a testable hypothesis, at least in a coarse sense.  
Figure~\ref{fig_ecc}a shows the normalized distributions of axial ratios for
12,045 spiral and 1,022 elliptical galaxies in the Third Reference Catalog of
Bright Galaxies \citep[RC3:][]{dev91}.  The axial ratios of the unclassified 2MASS 
QSO hosts are shown in Figure~\ref{fig_ecc}b.  There is a $\sim3\%$ and $\sim63\%$ 
probability that the unclassified host axial ratios are drawn from the
distributions of the RC3 elliptical and spiral galaxies, respectively, as 
indicated by KS tests.  Both of these results are inconclusive, except to 
say that it is more likely that the unclassified hosts are all spirals than
that they are all ellipticals.   Their nature can be investigated more generally
with a Monte Carlo approach. A total of $N$ spiral axial ratios and $12-N$ elliptical 
axial ratios were selected at random using the RC3 distributions as parent 
populations.  These 16 values were then binned to the same degree as 
Figure~\ref{fig_ecc}.  Repeating this exercise 10,000 times resulted in an 
average distribution for $N$ spiral galaxies as well as a Gaussian error for
each bin.  Computing $\chi^2$ indicated how well the unclassified
2MASS QSO hosts can be described as a sample of $N$ spiral galaxies and 
remaining ellipticals.  This test was
not very sensitive to the precise choice of N, but it showed that the unclassified
hosts can be characterized equally well as being purely spiral galaxies
or evenly divided between the two morphologies.

The preceding analysis begs the question of whether or not the 2MASS QSOs are exclusively found
in spiral galaxies.  For this to be the case, the four hosts classified as 
ellipticals would in fact have to be bulge-dominated spiral galaxies for which 
the disk was simply missed.  However, two of these lie at fairly low
redshifts ($z=0.140$ and $z=0.199$) where disks would likely be detected
if present.  
Also, two-thirds of the 2MASS QSOs with unclassified hosts also exhibit 
evidence of interaction including tidal debris, obviously merging galaxies and 
general asymmetry.  This is true of only two other hosts, one spiral and one 
elliptical.  Thus, what is seen clearly at low redshift as two closely separated
galaxies, might instead appear to be a nearly edge-on
spiral galaxy when seen at a higher redshift (e.g., 2M034857+2155 and
2M010607+2603).  The results of this snapshot survey seem to indicate that 
2MASS QSOs are found in both spiral and elliptical host galaxies (possibly 
more often in the former) exhibiting a range of dynamical activity.

\vskip 12pt
\begin{figure}
\centering
\plotone{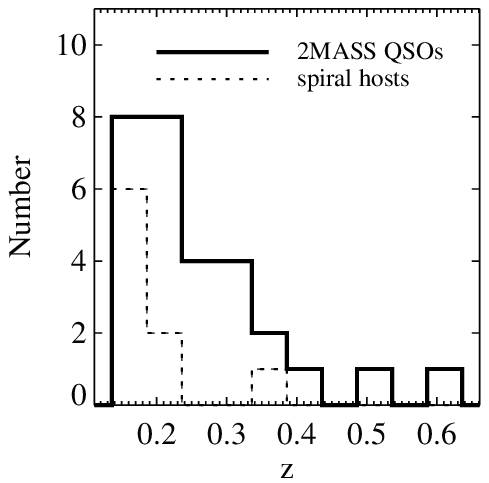}
\vskip 6pt
\figcaption{Redshift distributions of 2MASS QSOs investigated in this paper.  The dotted line 
indicates the subset of values for the host galaxies identified as spirals.}
\end{figure}

\subsection{Inclinations}

The distribution of inclinations for the disk galaxies in the sample
can potentially reveal selection effects caused by obscuring dust local
to the nucleus or in the host galaxy. Inclinations were calculated while 
accounting for the intrinsic axial ratio, $\alpha$, of an edge-on galaxy using
\begin{equation}
\cos i=\sqrt{\frac{q^{2}-\alpha ^{2}}{1-\alpha ^{2}}}\mathrm{,}
\end{equation}
where $q$ is the axial ratio b/a and $\alpha =0.2$
\citep{hub26,hol58}. The inclinations for each host are listed in Table~\ref{tbl_properties}
and their distribution is shown in Figure~\ref{fig_inclinations}. The mean inclination
is $67^{\circ}$, with only one inclination
greater than $75^{\circ}$ and none less than $50^{\circ}$. 

Despite our limited sample size (nine), there is an apparent disparity between
the inclinations of the 2MASS QSO hosts and those of UV-selected AGN.  With the exception
of hard X-ray selected samples, spirals hosting AGN preferentially have

\begin{table*}[h]
\begin{center}
\vspace*{12pt}
\tablenum{2}
{\sc Table \ref{tbl_properties}\\
\smallskip\hbox to\hsize{\hfil{Measured properties of 2MASS QSOs}\hfil}}
\small
\setlength{\tabcolsep}{6pt}
\begin{tabular}{ccccccccc}
\tskip \tableline \tskip
\colhead{Object} & \multicolumn{2}{c}{Nucleus\tablenotemark{a}} & 
\colhead{} & \multicolumn{5}{c}{Host Galaxy\tablenotemark{a}} \\
\cline{2-3}
\cline{5-9}
\colhead{(2MASSI J)} & \colhead{F814W\tablenotemark{b}} & \colhead{$M_I$} & 
\colhead{} & \colhead{$M_I$} & \colhead{Morphology} & \colhead{$b/a$} & 
\colhead{$i$} & \colhead{Interaction?\tablenotemark{c}} \\
\tskip \tableline
\tableline \tskip
000810.8+135452 &  $23.5$ & $-17.0$ & & $-22.6$ &     spiral &    0.61 & 54\degr &                    N\\
005055.7+293328 &  $19.5$ & $-20.3$ & & $-22.3$ &     spiral &    0.46 & 65\degr &                    N\\
010607.7+260334 &  $19.8$ & $-22.5$ & & $-20.9$ &          ? &    0.21 & \nodata &                    N\\
015721.0+171248 &  $20.1$ & $-20.7$ & & $-21.9$ &          ? &    0.68 & \nodata &                    Y\\
022150.6+132741 &  $20.6$ & $-19.3$ & & $-22.9$ & elliptical &    0.68 & \nodata &                    Y\\
023430.6+243835 &  $17.8$ & $-23.9$ & & $-23.1$ &    \nodata & \nodata & \nodata &                    N\\
032458.2+174849 &  $17.6$ & $-24.2$ & & $-23.9$ &          ? &    0.87 & \nodata &                    N\\
034857.6+125547 &  $21.4$ & $-19.4$ & & $-22.1$ &          ? &    0.41 & \nodata &                    Y\\
092049.0+190320 &  $21.0$ & $-19.1$ & & $-21.1$ &     spiral &    0.39 & 70\degr &                    N\\
125807.4+232921 &  $17.7$ & $-23.6$ & & $-22.3$ &          ? &    0.41 & \nodata &                    N\\
130700.6+233805 &  $22.6$ & $-18.8$ & & $-22.7$ & elliptical &    0.81 & \nodata &                    N\\
145331.5+135358 &  $20.4$ & $-19.5$ & & $-22.7$ &     spiral &    0.48 & 64\degr &                    N\\
151621.1+225944 &  $18.9$ & $-21.6$ & & $-22.1$ &     spiral &    0.37 & 72\degr &                    N\\
152151.0+225120 &  $21.9$ & $-19.5$ & & $-23.1$ &          ? &    0.57 & \nodata &                    Y\\
154307.7+193751 &  $18.1$ & $-22.8$ & & $-23.9$ &          ? &    0.74 & \nodata &                    Y\\
163700.2+222114 &  $19.3$ & $-21.5$ & & $-23.0$ &     spiral &    0.27 & 80\degr &                    Y\\
163736.5+254302 &  $20.6$ & $-20.8$ & & $-22.9$ &          ? &    0.86 & \nodata &                    Y\\
165939.7+183436 &  $18.0$ & $-22.3$ & & $-22.7$ &          ? &    0.44 & \nodata &                    Y\\
170003.0+211823 &  $19.7$ & $-23.5$ & & $-22.5$ &    \nodata & \nodata & \nodata &                    N\\
171442.7+260248 &  $17.9$ & $-22.3$ & & $-22.7$ &     spiral &    0.52 & 61\degr &                    N\\
171559.7+280717 &  $20.7$ & $-22.1$ & & $-23.9$ &          ? &    0.65 & \nodata &                    N\\
222202.2+195231\tablenotemark{d} &  $18.9$ & $-23.1$ & & \nodata &    \nodata & \nodata & \nodata &                    N\\
222221.1+195947 &  $17.5$ & $-23.2$ & & $-22.6$ &          ? &    0.77 & \nodata &                    N\\
222554.2+195837 &  $22.2$ & $-17.8$ & & $-23.2$ &     spiral &    0.39 & 70\degr &                    N\\
225902.5+124646 &  $18.9$ & $-21.8$ & & $-22.8$ & elliptical &    0.57 & \nodata &                    N\\
230304.3+162440 &  $23.1$ & $-18.4$ & & $-22.7$ &          ? &    0.29 & \nodata &                    Y\\
230442.4+270616 &  $20.5$ & $-20.5$ & & $-22.5$ & elliptical &    0.74 & \nodata &                    N\\
232745.6+162434 &  $22.1$ & $-19.9$ & & $-23.8$ &     spiral &    0.49 & 63\degr &                    N\\
234449.5+122143 &  $17.3$ & $-23.3$ & & $-23.0$ &    \nodata & \nodata & \nodata &                    Y\\
\tskip\tableline
\end{tabular}
\vspace*{-12pt}
\tablenotetext{a}{Nuclear and host magnitudes are the result of PSF-subtraction within $10\arcsec$ apertures.  The total uncertainty in this decomposition can be as large as one magnitude.}
\tablenotetext{b}{Apparent magnitude in \emph{HST} F814W filter (not K-corrected).}
\tablenotetext{c}{Double-nuclei, merging galaxies or disturbed morphologies such as tidal debris and obvious asymmetry.}
\tablenotetext{d}{The sample PSF was created from 2MASSI J222202.2+195231.}
\end{center}
\end{table*}

\begin{figure*}[b]
\epsscale{1.56}
\vskip 12pt
\begin{center}
\plottwo{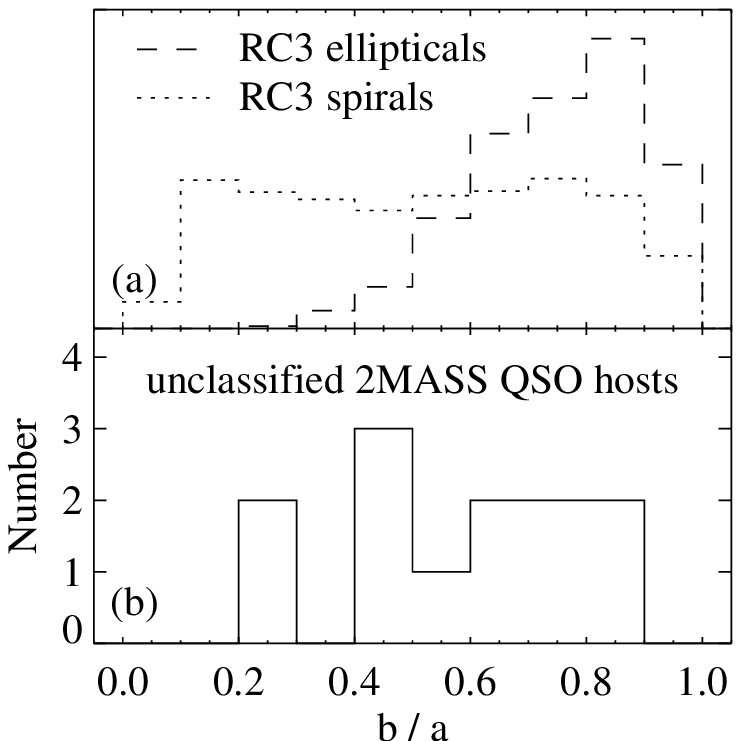}{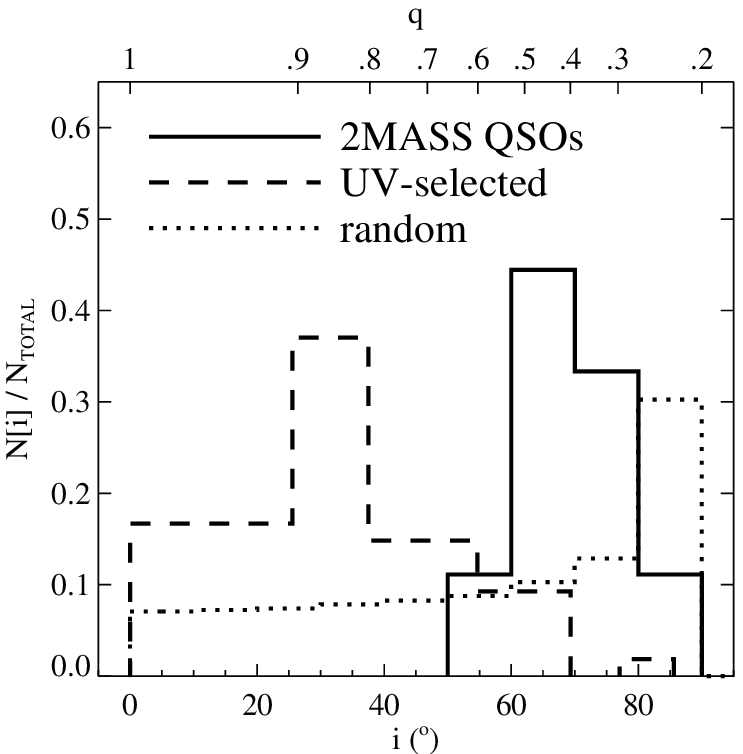}
\vskip 6pt
\caption{\emph{left} -- (a) Distributions of axial ratios for 12,045 spiral 
and 1,022 elliptical galaxies in the RC3 Catalog \citep{dev91}.
(b) Measured axial ratios of 2MASS QSO host galaxies with unclassified morphologies.}
\caption{\emph{right} -- Spiral host galaxly inclinations of 2MASS QSOs and UV-selected AGN \citep{sim97}, as well
as the expectation for randomly oriented disks. The area under the curves
have been normalized to one. Therefore, the vertical axis represents the 
fractional number of inclinations in each bin.}
\end{center}
\end{figure*}

\clearpage
\noindent low inclinations 
\citep{sim97}. \citet{mai95} found a more even distribution of inclinations in a sample
of very nearby Seyferts ($\langle d \rangle=34$ Mpc) in which nuclei with low
apparent brightness were observable.
They suggested that a 100 pc-scale torus coplanar with the galactic disk
could be responsible for obscuring nuclear activity in highly-inclined disk galaxies.  
This is consistent with the results of our study.  Reddened AGN in highly-inclined 
spirals that are missed by UV/optical surveys could be detected in the near-infrared.
Furthermore, low inclinations would not be found in our sample because a less obstructed 
view of the nucleus would not satisfy the 2MASS red AGN color criterion.

A lack of the highest inclinations is not surprising.
At $i=90^{\circ}$ the nucleus would be viewed through the entire plane of the 
galaxy and not likely seen at all.  Also, the 2MASS selection criteria 
require objects to be seen in all three wavebands.  Extremely red 
``J-band drop-outs'' that were excluded from the original 2MASS AGN sample, 
may include objects with host galaxies seen nearly edge-on.

\begin{table}[b]
\begin{center}
\tablenum{3}
{\sc Table \ref{tbl_companions}\\
\smallskip\hbox to\hsize{\hfil{2MASS QSO \emph{apparent} companions}\hfil}}
\small
\setlength{\tabcolsep}{6pt}
\begin{tabular}{cccrc}
\tskip \tableline \tskip
\colhead{Object} & \colhead{$M_{I}$} & \multicolumn{3}{c}{projection} \\
\colhead{(2MASSI J)} & \colhead{} & \multicolumn{3}{c}{(h$^{-1}$ kpc)} \\
\tskip \tableline
\tableline \tskip
015721.0+171248 & $-18.4$ & &	5.9 &  \\
034857.6+125547 & $-20.5$ & &	3.5 &  \\
092049.0+190320 & $-19.5$ & &	6.4 &  \\
152151.0+225120 & $-19.3$ & &	4.1 &  \\
154307.7+193751 & $-21.5$ & &  14.4 &  \\
163736.5+254302 & $-20.2$ & &	1.2 &  \\
165939.7+183436 & $-20.9$ & &	1.2 &  \\
171442.7+260248 & $-18.8$ & &  17.0 &  \\
		& $-19.3$ & &  14.9 &  \\
                & $-19.0$ & &  12.4 &  \\
222221.1+195947 & $-19.2$ & &	7.9 &  \\
225902.5+124646 & $-19.1$ & &  10.8 &  \\
230304.3+162440 & $-22.4$ & &	3.0 &  \\
230442.4+270616 & $-19.8$ & &	5.0 &  \\
	        & $-19.1$ & &  16.0 &  \\
234449.5+122143 & $-21.8$ & &	2.9 &  \\
\tskip\tableline
\end{tabular}
\end{center}
\end{table}

\begin{table}[t]
\begin{center}
\tablenum{3}
{\sc Table \ref{tbl_companions_ulirgs}\\
\smallskip\hbox to\hsize{\hfil{QDOT ULIRG \emph{apparent} companions}\hfil}}
\small
\setlength{\tabcolsep}{6pt}
\begin{tabular}{ccccrc}
\tskip \tableline \tskip
\colhead{Object} & \colhead{z} & \colhead{$M_{I}$} & \multicolumn{3}{c}{projection} \\
\colhead{(IRAS)} & \colhead{} & \colhead{} & \multicolumn{3}{c}{(h$^{-1}$ kpc)} \\
\tskip \tableline
\tableline \tskip
00207+1029   & 0.230  & $-19.3$  & &  11.9  & \\
	       &        & $-22.2$  & &   1.8  & \\
00275$-$2859 & 0.280  & $-19.8$  & &   3.6  & \\
	       &        & $-19.2$  & &   7.4  & \\
00461$-$0728 & 0.243  & $-21.0$  & &   1.5  & \\
01185+2547   & 0.184  & $-21.0$  & &   2.7  & \\
01284$-$1535 & 0.153  & $-20.2$  & &   1.7  & \\
02054+0835   & 0.345  & $-18.5$  & &  19.1  & \\
	       &        & $-19.3$  & &  12.2  & \\
	       &        & $-20.8$  & &  15.7  & \\
	       &        & $-19.6$  & &  12.7  & \\
	       &        & $-18.9$  & &   5.9  & \\
02587$-$6336 & 0.255  & $-18.8$  & &  21.0  & \\
	       &        & $-21.5$  & &  13.1  & \\
03538$-$6432 & 0.301  & $-22.8$  & &   1.6  & \\
04384$-$4848 & 0.204  & $-20.6$  & &   1.4  & \\
05120$-$4811 & 0.163  & $-19.0$  & &  14.4  & \\
	       &        & $-18.9$  & &  17.1  & \\
	       &        & $-20.6$  & &   1.8  & \\
08201+2801   & 0.168  & $-21.0$  & &   2.2  & \\
10026+4347   & 0.178  & $-18.9$  & &  18.9  & \\
10558+3845   & 0.207  & \nodata  & & \nodata & \\ 
10579+0438   & 0.173  & $-20.0$  & &   2.6  & \\
12108+3157   & 0.206  & $-20.4$  & &   0.9  & \\
12202+1646   & 0.181  & $-20.9$  & &  21.6  & \\
13469+5833   & 0.158  & $-19.3$  & &   4.7  & \\
	       &        & $-21.5$  & &   2.9  & \\
15168+0045   & 0.154  & $-19.0$  & &  17.3  & \\
	       &        & $-22.0$  & &  16.2  & \\
16159$-$0402 & 0.213  & $-20.9$  & &   6.3  & \\
16455+4553   & 0.191  & $-19.8$  & &   0.5  & \\
165411+530   & 0.194  & $-22.1$  & &   4.0  & \\
171754+544   & 0.147  & $-19.0$  & &  24.0  & \\
	       &        & $-21.2$  & &   0.6  & \\
17469+5806   & 0.309  & $-18.5$  & &  22.9  & \\
	       &        & $-19.5$  & &  17.9  & \\
	       &        & $-20.9$  & &   1.1  & \\
20176$-$4756 & 0.178  & $-19.0$  & &   5.6  & \\
20253$-$3757 & 0.180  & $-21.3$  & &   4.2  & \\
	       &        & $-21.0$  & &   2.1  & \\
20314$-$1919 & 0.153  & $-20.3$  & &  23.6  & \\
	       &        & $-20.5$  & &   3.9  & \\
20507$-$5412 & 0.228  & $-20.7$  & &   1.1  & \\
21547$-$5823 & 0.165  & $-22.3$  & &  10.2  & \\
23140+0348   & 0.220  & $-18.9$  & &  15.4  & \\
23498+2423   & 0.212  & $-21.0$  & &   8.5  & \\
\tskip\tableline
\end{tabular}
\vspace*{-12pt}
\end{center}
\end{table}

\subsection{Apparent Companions}\label{sec_env}

Previous studies have shown that low-redshift QSOs generally lie in 
moderate galaxy groups as opposed to rich clusters or field densities 
\citep[and references therein]{har90,bah91,fis96}.  The environment in 
which this sample of 2MASS-selected QSOs is found was characterized
by studying the frequency and distribution of \emph{apparent} companion 
galaxies.  This refers to all extended objects with
small projected separations.  Redshift and color information are 
unavailable and only magnitude considerations can be used to eliminate
some background galaxies.  Clearly, the number of apparent companions
is greater than or equal to the number of true companions that, by
definition, lie in close proximity to the QSOs.
The presence of physically unrelated objects will also 
dilute any non-random radial distribution (in the plane of the sky) 
of companion galaxies.  However, for the
purpose of comparative studies, background galaxies do not pose a systematic 
problem as long as the same identification criteria are employed for all
studies.

Apparent companions were selected in the following way.  The images were 
marginally smoothed and high-density regions with surface 
brightness exceeding the image noise by $\ge3\sigma$ were flagged.  The
FWHM was then measured for each candidate
and compared to that of the previously constructed PSF as well as 
with obvious field stars.  Objects with a FWHM less than the adopted two pixel
cutoff, as well as those possessing diffraction spikes in the case of
stars with saturated cores, were rejected.  Finally, an absolute 
magnitude criterion ($M_I \leq 
-18.3$) was imposed assuming the same redshift as the QSO.  This 
serves to discriminate against more luminous objects at larger, unrelated distances.

\begin{figure*}[t]
\epsscale{.35}
\vskip 12pt
\hskip 110pt
\plotone{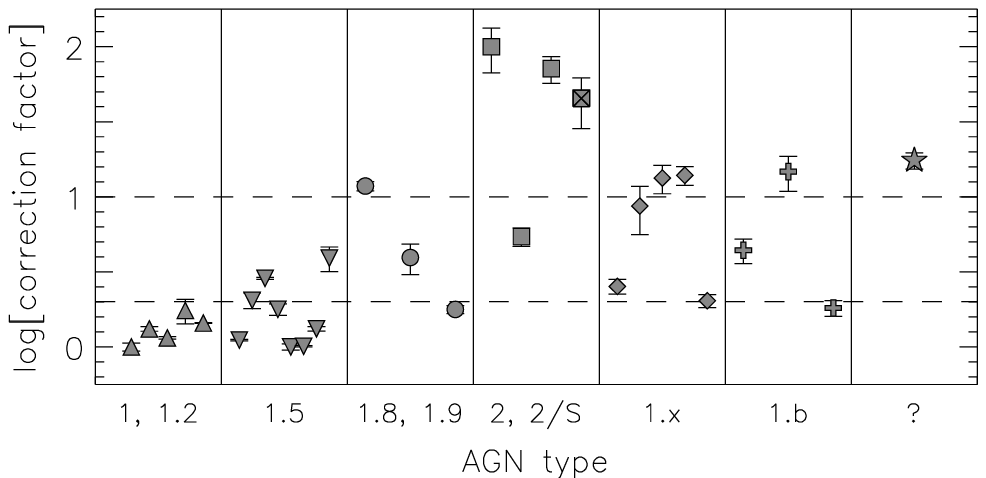}
\vskip 6pt
\caption{Polarization corrections and $3\sigma$ error bars for all 29 AGN due to dilution from host galaxy starlight.  Dashed lines indicate correction factors of two and ten.  Each symbol
corresponds to an AGN spectral type bin (the $\boxtimes$ refers to the 2/S object).  The corrections correlate with AGN type (type 1 through 2).}
\end{figure*}

The identical procedure was performed on the sample of 30 ULIRGs
drawn from the \emph{HST} archive.
To account for the different pixel scale of the WF3 
camera, these images were first resampled to match the 2MASS QSO 
images observed with the PC.  This allowed us to select the high
density surface brightness regions with the same criteria, preventing
any bias that might have resulted from different angular pixel sizes.  
The increased sensitivity 
of the wide field cameras to low surface brightness should not
present a problem since the magnitude cutoff was substantially brighter
than the limiting magnitudes of the 2MASS QSO images.

Apparent companions found within 25 $h^{-1}$ projected kpc of the 
2MASS QSOs are presented in Table~\ref{tbl_companions}.  
Table~\ref{tbl_companions_ulirgs} provides complimentary information
for the QDOT ULIRGs in addition to their redshifts.
Further discussion of these
companions is deferred to \S \ref{sec_discussion} where a third
sample will be considered.

\subsection{Polarization}

The optical broadband linear polarization of 70 2MASS QSOs (including all 29 
objects addressed by this paper) has been measured, and a detailed 
discussion of their properties can be found in \citet{smi02}.
More than 10\% of 2MASS QSOs show high polarization ($P>3\%$), ranging up
to $~11\%$.  Lower polarization values were generally observed for lower 
luminosity AGN, a trend that can be understood in part by dilution of 
polarized light by unpolarized light from the host galaxy.

The PSF fitting in \S \ref{sec_psf} provides a measure of 
the polarization dilution via starlight. If the linear polarization, $P$, is measured with 
an uncertainty $\sigma_P$, and the total flux, $F_{TOT}$, has nuclear contribution 
$F_{AGN}$ with uncertainty $\sigma_{F_{AGN}}$, then the 
nuclear polarization, $P_{AGN}$, and uncertainty, $\sigma_{P_{AGN}}$, are given by 
\begin{eqnarray}
P_{AGN} & = & P \left( \frac{F_{TOT}}{F_{AGN}} \right) \\
\sigma_{P_{AGN}} & = & \frac{F_{TOT}}{F_{AGN}} 
    \left[ \sigma_P^2+\sigma_{F_{AGN}}^2 
    \left( \frac{P}{F_{AGN}} \right) ^2 \right] ^{1/2}.
\end{eqnarray}
Columns 2 and 4 of Table~\ref{tbl_pol} list the measured and derived values.
No adjustment for statistical bias has been made.

\begin{table}[b]
\vspace*{-6pt}
\begin{center}
{\sc Table \ref{tbl_pol}\\
\smallskip\hbox to\hsize{\hfil{Polarization corrections for starlight dilution}\hfil}}
\small
\setlength{\tabcolsep}{6pt}
\begin{tabular}{l r@{$\pm$}l r@{$\pm$}l r@{$\pm$}l}
\tskip \tableline \tskip
\colhead{Object} & \multicolumn{2}{c}{$P$\tablenotemark{a}} & 
\multicolumn{2}{c}{$f\frac{AGN}{TOT}$\tablenotemark{b}} & \multicolumn{2}{c}{$P_{AGN}$} \\
\colhead{(2MASSI J)} & \multicolumn{2}{c}{$(\%)$} & \multicolumn{2}{c}{} & \multicolumn{2}{c}{$(\%)$} \\
\tskip \tableline
\tableline \tskip
000810.8+135452 &  0.79&0.47 &0.01&0.001 &78.83&47.70\\
005055.7+293328 &  2.47&0.49 &0.18&0.009 &13.42& 2.74\\
010607.7+260334 &  6.03&0.54 &1.00&0.020 & 6.03& 0.55\\
015721.0+171248 &  1.44&0.47 &0.40&0.015 & 3.64& 1.20\\
022150.6+132741 &  0.39&0.24 &0.08&0.002 & 4.60& 2.83\\
023430.6+243835 &  2.57&0.46 &0.90&0.003 & 2.85& 0.51\\
032458.2+174849 &  1.23&0.38 &0.76&0.009 & 1.62& 0.50\\
034857.6+125547 &  2.23&0.73 &0.12&0.014 &19.36& 6.74\\
092049.0+190320 &  1.15&0.52 &0.23&0.014 & 5.07& 2.31\\
125807.4+232921 &  1.22&0.13 &0.87&0.006 & 1.40& 0.15\\
130700.6+233805 &  2.45&0.63 &0.07&0.006 &36.10& 9.80\\
145331.5+135358 &  0.81&0.29 &0.07&0.005 &10.80& 3.94\\
151621.1+225944 &  1.02&0.28 &0.55&0.022 & 1.85& 0.51\\
152151.0+225120 &  1.02&0.48 &0.07&0.003 &14.18& 6.71\\
154307.7+193751 &  1.33&0.26 &0.49&0.020 & 2.72& 0.54\\
163700.2+222114 &  2.34&0.40 &0.35&0.002 & 6.70& 1.15\\
163736.5+254302 &  0.42&0.28 &0.25&0.019 & 1.65& 1.11\\
165939.7+183436 &  6.30&0.73 &0.56&0.016 &11.21& 1.34\\
170003.0+211823 & 11.11&0.80 &1.00&0.015 &11.11& 0.82\\
171442.7+260248 &  0.86&0.33 &0.57&0.036 & 1.50& 0.58\\
171559.7+280717 &  6.08&1.28 &0.56&0.011 &10.81& 2.29\\
222202.2+195231 &  7.19&1.14 &0.99&0.003 & 7.26& 1.15\\
222221.1+195947 &  0.95&0.23 &0.76&0.009 & 1.25& 0.30\\
222554.2+195837 &  1.38&0.56 &0.01&0.001 &98.82&40.65\\
225902.5+124646 &  1.03&0.78 &0.49&0.016 & 2.09& 1.59\\
230304.3+162440 &  0.24&0.51 &0.02&0.003 &10.85&23.09\\
230442.4+270616 &  0.24&0.55 &0.26&0.016 & 0.94& 2.15\\
232745.6+162434 &  1.08&0.57 &0.06&0.002 &18.89&10.00\\
234449.5+122143 &  1.01&0.24 &0.69&0.002 & 1.46& 0.35\\
\tskip \tableline
\end{tabular}
\vspace*{-24pt}
\tablenotetext{a}{Linear polarization measurements (not corrected for statistical bias) from \citet{smi02}.}
\tablenotetext{b}{Fractional contribution of AGN to total light (within $10\arcsec$ radius aperture) in HST filter F814W.}
\end{center}
\end{table}

Figure~\ref{fig_pol3} shows the range of polarization
corrections ($=F_{TOT}/F_{AGN}$) with $3\sigma$ error bars 
for the sample of 2MASS QSOs.  
The error-weighted mean correction is 1.46, although 52\% 
(15/29) of the corrections are \emph{at least} a factor of two, 
and 28\% (8/29) are ten or greater even
with consideration of the margin of error.  Clearly, the flux contribution from host
galaxies represents an important correction for the polarization 
of many 2MASS QSOs.
Without this correction, 17\% (5/29) 
of the sample had measured
polarizations in excess of $3\%$, all significant 
at greater than the $4\sigma$ level.  
With the correction, 62\% (18/29) have $P>3\%$ though only 24\% 
(7/29) have both $P>3\%$
and $P/\sigma_P>4$.  The AGN is likely more obscured in the $R$-band 
(approximately where the polarization 
measurements were made) than in the $I$-band (where the starlight dilution was measured), so 
the correction factors may be even larger than stated.

The 2MASS QSOs reflect the full range of AGN types from type 1 to type 2 (Table~\ref{tbl_sample}).
Using the spectral classification adopted by \citet{cut01}, type ``1.x'' refers to objects 
that exhibit broad $H\alpha$ emission lines but have no detected 
$H\beta$. Objects that show 
broad $H\alpha$ but whose spectral coverage does not include $H\beta$ are labeled ``1.b''.  
Classification of 2M232745+1624 is ambiguous. A relationship is evident between AGN type and the 
degree of dilution due to 
host galaxy starlight (Figure~\ref{fig_pol3}). Type 1 and 1.2 AGN have the smallest 
$F_{TOT}/F_{AGN}$ ratios while type 1.8-2 are dominated by the host 
galaxy.  This result is consistent with the assertion by \citet{smi02} and the findings
of \citet{sch02} and \citet{smi03} that the low polarization of 
type 1.8-2 objects is largely caused by the contribution of 
unpolarized starlight.

\section{Comparison Studies}\label{sec_compsamp}

The aim of this snapshot survey was to characterize the host galaxies and 
environments in which the red 2MASS QSOs are found.
Directly comparing these properties to those of previously investigated samples
offers insight into the nature of the 2MASS QSOs and allows us to address
the questions raised in \S \ref{sec_intro}.  Three samples of ULIRGs were 
selected from the literature, as well as five studies of QSOs and their 
host galaxies.  They were primarily chosen on grounds of suitability for useful 
comparison and are, therefore, a representative selection rather than an 
exhaustive one.

\emph{\citet{sur98}} used \emph{HST} to study nine ULIRGs, employing the same 
instrumental setup as this 2MASS QSO snapshot survey (WFPC2/PC/F814W).  Color
information allowed these authors to investigate a number of issues which are
beyond the scope of discussion in this paper.  However, they do report the 
incidence of interaction among their sample.

\emph{\citet{far01}} imaged 23 QDOT ULIRGs with \emph{HST}'s F606W filter
($\sim$\emph{V}-band) and the PC chip of WFPC2.
They present observed properties including morphology and interaction.
Only twelve of their objects have $|b|>30^{\circ}$,
a necessary condition for fairly comparing their surrounding environoments 
to those of the 2MASS QSOs.  For this reason, apparent companions were identified for 
a different (but overlapping) sample of ULIRGs as was discussed in 
\S \ref{sec_ulirgs}.

\emph{\citet{kim02}} recently obtained ground-based optical ($R$-band) and 
near-infrared ($K'$-band) imaging of the \emph{IRAS} 1-Jy sample of 118 ULIRGs.
The analysis of these images is presented in \citet{vei02}.  Among other things,
these authors address morphology and the incidence of interaction.

The five QSO samples were constructed with varying selection criteria and
include significant overlap.  To ensure a homogeneous data set, only the 
radio-quiet PG QSOs were considered.  We briefly describe each study before
discussing the aggregate PG QSO sample.

\emph{\citet{mcl01}} observed sixteen radio-quiet QSOs in the $H$-band 
with the Near-Infrared Camera and Multi-Object Spectrometer (NICMOS) on 
\emph{HST}.  Ten of these objects were originally part of
a sample of 26 high-luminosity 
QSOs ($M_B<-23.1$ and $z\leq0.3$) selected from the Bright Quasar Survey 
\citep{sch83} and studied by \citet{mcl94b}.  The additional six objects
were drawn from a slightly larger redshift range ($z<0.4$).  

\emph{\citet{per01}} obtained deep $K$-band ground-based imaging for a sample of 14 high 
luminosity, radio-quiet QSOs.  The observations were made with the 
UK Infrared Telescope (UKIRT) equipped with the IRCAM3 camera.
Their sample has total absolute magnitudes of $M_V < -25$ and
span the redshift range $0.26 < z < 0.46$.

\emph{\citet{sur01}} carried out a ground-based optical and near-infrared study of 
17 ``infrared-excess'' PG QSOs (all PG QSOs with $z\le0.16$ and 
$L_{ir}/L_{BBB}$\footnote{As described in \citet{sur01}, $L_{ir}=L(8\mathrm{-}1000 \micron)$ 
and $L_{BBB}$ is the ``big blue bump'' luminosity at 3200-8400 \AA.}$ > 0.46$)
in order to investigate whether 
ULIRGs represent the progenitors of optically selected QSOs 
\citep{san88}.  With regard to the properties addressed in our comparison
study, these PG QSOs are indistinguishable from the others and respresent
a significant contribution to host galaxy studies of radio-quiet PG QSOs.

\emph{\citet{mclu99}} conducted an optical imaging study of six radio-loud QSOs,
nine radio-quiet QSOs and four radio galaxies.
These objects were selected from a slightly larger sample ($0.1<z<0.35$) 
previously imaged in the 
near-IR \citep{dun93}.  Observations were made with \emph{HST}/WFPC2, and targets 
were centered on the Wide-Field 2 (WF2) chip.  The F675W filter, which is roughly equivalent to 
the standard $R$-band, was used.  

\emph{\citet{bah97}} pointed \emph{HST} at the 14 most luminous QSOs ($z<0.2$) in 
the \citet{ver91} catalogue as well as six additional objects ($0.2<z<0.3$) 
satisfying the same luminosity and galactic latitude criteria ($M_V<-22.9$ 
and $|b|>35^{\circ}$).  Observations were centered on the WF3 camera of 
WFPC2 and used filter F606W.  In addition to addressing host morphology
and luminosity, the authors carried out a simple analysis of the host environment 
by identifying
all apparent companions\footnote{``Companions'' were defined as extended objects with 
$M_V \le -16.5$, assuming the same redshift as the host galaxy.} within 25 projected kpc.
Magnitudes and separations are presented for 13 such objects.  

As a whole, these five studies contain 38 radio-quiet PG QSOs (Table~\ref{tbl_rqq}).  
In those cases where an object is included in multiple samples, the measured
luminosities and morphologies are generally in good agreement.  There are three 
ocurrences where a host galaxy was called an elliptical by one study and given
no morphological classification by another.  In two cases, investigators 
disagree whether a host is a spiral or an elliptical.  Thus, between 17 and 19 of
the radio-quiet PG QSOs are reported to lie in elliptical galaxies, 13 to 15 are
found in spirals and six hosts could not be classified.

The redshift distribution of this PG sample is comparable to that of the 2MASS
QSOs.  However, observations were made at a variety of wavelengths, and differing
cosmologies were considered.  Before comparisons could be made, luminosities 
first had to be translated to a common frame
of reference.  The $I$-band was chosen because it corresponds 
closely to the F814W filter, minimizing the degree of alteration to the 2MASS
QSO data.  This is especially important for the nuclear components, as their
SED is not well known.  

Absolute $I$-band magnitudes were calculated for the various samples
by first accounting for differing $H_0$ and $q_0$ values using a form 
of Equation~\ref{eq_abs}.  If primed values denote the originally preferred 
cosmology, and absolute magnitudes are given in the

\clearpage
\begin{table}[t]
\begin{center}
\vskip 12pt
{\sc Table \ref{tbl_rqq}\\
\smallskip\hbox to\hsize{\hfil{PG QSOs from literature}\hfil}}
\small
\setlength{\tabcolsep}{6pt}
\begin{tabular}{c c c c c c}
\tskip \tableline \tskip
\colhead{Object} & \colhead{z} & \colhead{Host} & \colhead{$M_{I_{HOST}}$} & 
\colhead{$M_{I_{NUC}}$} & \colhead{Inter-} \\
\colhead{(PG)} & \colhead{} & \colhead{} & \colhead{} & \colhead{} & \colhead{action?\tablenotemark{*}} \\
\tskip \tableline
\tableline \tskip
 0026+129     & 0.142 & E$^\mathrm{a}$ & $-$22.0 & $-$23.9 & N \\
 0043+039     & 0.384 & S$^\mathrm{b}$ & $-$22.7 & $-$24.1 & \nodata \\
 0052+251     & 0.155 & S$^\mathrm{e}$ & $-$23.1 & $-$23.7 & N \\
 0054+144     & 0.171 & E$^\mathrm{d}$ & $-$23.5 & $-$23.8 & N \\
		&       & E$^\mathrm{e}$ & $-$23.4 & $-$23.5 & N \\
 0157+001     & 0.163 & ?$^\mathrm{d}$ & $-$24.1 & $-$23.0 & Y \\
		      &	    & E$^\mathrm{c}$ & \nodata & \nodata & \nodata \\
 0838+770     & 0.131 & S$^\mathrm{c}$ & $-$22.6 & $-$23.0 & N \\
 0923+201     & 0.190 & E$^\mathrm{d}$ & $-$23.1 & $-$23.9 & N \\
		      &       & E$^\mathrm{e}$ & $-$23.1 & $-$23.8 & N \\
 0947+396     & 0.206 & S$^\mathrm{a}$ & $-$22.5 & $-$23.3 & N \\
 0953+415     & 0.239 & E$^\mathrm{d}$ & $-$22.7 & $-$24.8 & N \\
		      &	    & ?$^\mathrm{e}$ & $-$22.3 & $-$25.0 & Y \\
 1001+054     & 0.161 & ?$^\mathrm{c}$ & $-$21.0 & $-$23.5 & N \\
 1012+008     & 0.185 & E$^\mathrm{d}$ & $-$23.6 & $-$23.2 & Y \\
		      &	    & ?$^\mathrm{e}$ & $-$23.9 & $-$23.6 & Y \\
 1029$-$140   & 0.086 & E$^\mathrm{e}$ & $-$22.7 & $-$24.0 & N \\
 1048+342     & 0.167 & E$^\mathrm{a}$ & $-$22.5 & $-$23.4 & Y \\
 1114+445     & 0.144 & S$^\mathrm{c}$ & $-$21.6 & $-$24.0 & Y \\
 1116+215     & 0.177 & E$^\mathrm{e}$ & $-$23.6 & $-$24.8 & N \\
 1121+422     & 0.224 & E$^\mathrm{a}$ & $-$21.2 & $-$23.4 & N \\
 1126$-$041   & 0.060 & S$^\mathrm{c}$ & $-$22.3 & $-$22.9 & Y \\
 1151+117     & 0.176 & ?$^\mathrm{a}$ & $-$21.7 & $-$23.8 & N \\
 1202+281     & 0.165 & E$^\mathrm{e}$ & $-$22.7 & $-$23.8 & N \\
		      &	    & E$^\mathrm{c}$ & \nodata & \nodata & \nodata \\
 1216+069     & 0.334 & E$^\mathrm{b}$ & $-$22.3 & $-$24.4 & \nodata \\
 1229+204     & 0.064 & S$^\mathrm{c}$ & $-$22.6 & $-$22.0 & Y \\
 1307+085     & 0.155 & E$^\mathrm{e}$ & $-$22.5 & $-$24.1 & N \\
 1309+355     & 0.184 & S$^\mathrm{e}$ & $-$23.1 & $-$23.9 & N \\
 1322+659	    & 0.168 & S$^\mathrm{a}$ & $-$21.8 & $-$23.4 & N \\
 1351+640     & 0.088 & ?$^\mathrm{c}$ & $-$22.4 & $-$23.9 & N \\
 1352+183     & 0.158 & E$^\mathrm{a}$ & $-$22.3 & $-$23.4 & N \\
 1354+213     & 0.300 & E$^\mathrm{a}$ & $-$22.8 & $-$24.7 & N \\
		      &	    & S$^\mathrm{b}$ & $-$22.3 & $-$23.2 & \nodata \\
 1402+261     & 0.164 & S$^\mathrm{c}$ & $-$21.7 & $-$24.1 & N \\
		      &  	    & S$^\mathrm{e}$ & $-$22.4 & $-$23.8 & N \\
 1411+442     & 0.090 & ?$^\mathrm{c}$ & $-$22.1 & $-$23.7 & Y \\
 1415+451     & 0.114 & ?$^\mathrm{c}$ & $-$22.1 & $-$23.7 & N \\
 1427+480     & 0.221 & E$^\mathrm{a}$ & $-$22.1 & $-$23.5 & Y \\
 1440+356     & 0.079 & S$^\mathrm{c}$ & $-$22.3 & $-$23.4 & N \\
 1444+407     & 0.267 & E$^\mathrm{e}$ & $-$22.8 & $-$24.7 & N \\
 1543+489     & 0.400 & S$^\mathrm{b}$ & $-$22.4 & $-$24.8 & \nodata \\
 1613+658     & 0.129 & E$^\mathrm{a}$ & $-$23.7 & $-$23.2 & Y \\
		      &	    & ?$^\mathrm{c}$ & $-$23.6 & $-$24.1 & Y \\
 2112+059     & 0.466 & ?$^\mathrm{b}$ & $-$22.1 & $-$25.4 & \nodata \\
 2130+099     & 0.062 & S$^\mathrm{c}$ & $-$22.5 & $-$23.0 & N \\
 2233+134     & 0.325 & E$^\mathrm{a}$ & $-$22.3 & $-$24.4 & N \\
		&	      & S$^\mathrm{b}$ & $-$21.2 & $-$23.7 & \nodata \\
\tskip\tableline
\end{tabular}
\vspace*{-12pt}
\tablenotetext{*}{Determined by authors of this paper based on published imaging.  The same criteria
applied to the 2MASS sample were used (double-nuclei, merging galaxies or disturbed morphologies 
such as tidal debris and obvious asymmetry).}
\tablenotetext{a}{\citet{mcl01}}
\tablenotetext{b}{\citet{per01}}
\tablenotetext{c}{\citet{sur01}}
\tablenotetext{d}{\citet{mclu99}}
\tablenotetext{e}{\citet{bah97}}
\end{center}
\end{table}

\noindent $X$-band, then,
\begin{equation}
M_X = M^{'}_{X} + 5 \log[h/h^{'}] + 1.086 (q_0-q_0^{'}) z.
\end{equation}
\noindent The conversion from $X$-band to $I$-band is then given by,
\begin{equation}
M_I = M_X - 2.5 \log
   \left[\left(f_I/f_X\right)\left(f_{X,0}/f_{I,0}\right)\right],
\end{equation}
where $f_0$ refers to the flux of an A0V star in the band of interest
\citep{bes79,cam85}.  Values for $f_I$ and $f_X$ were measured in the same
way as the K-corrections described in \S \ref{sec_lum}.  The SDSS composite
QSO spectrum only extends to the rest $I$-band and could not be used to 
transform the $K$-band nuclear luminosities in \citet{per01}.  The red-end 
spectral index of the composite SED was used in this case, 
$f_\nu \propto \nu^{-1.5}$.

The resulting host and nuclear $I$-band luminosities are provided in 
Table~\ref{tbl_rqq}.  Of the eight objects included in more than one 
sample, five of them agree within half of a magnitude for both the 
host and the nucleus.  Disagreement ($\le1.5$ magnitudes) among the remaining
three objects can be understood in part by disparate host galaxy morphologies
resulting in different fitted luminosities.  Adopting the first value 
listed for each object in Table~\ref{tbl_rqq}, the absolute $I$-band magnitudes 
of the radio-quiet PG host galaxies range from 0.3 L* to 5.2 L* with a mean 
of 1.5 L*.  The corresponding nuclear values are $-21.7<M_I<-25.4$
and $\langle M_I \rangle=-23.7$.

\section{Discussion}\label{sec_discussion}

The galaxies in which the 29 2MASS QSOs are found do not differ dramatically
from those hosting classical PG QSOs.  Imaging available to date suggests that 
the latter are both ellipticals and spirals in roughly equal proportion.  The 
2MASS QSO hosts also include both morphological types, though there is some 
evidence that they are spirals more often than not.  As many as 26\% of the 
radio-quiet PG QSO hosts were not assigned a morphology by at least one 
investigating team. The larger fraction of the 
2MASS QSO hosts with ambiguous morphologies can be explained in part by the 
greater efforts invested by some of the comparison studies in modeling the
host and assigning a morphology.

The 2MASS QSOs are not found in comparatively underluminous or unusually 
bright galaxies.  The top panel of Figure~\ref{fig_agn_vs_gal} plots
nuclear versus host galaxy luminosity for the 2MASS and PG QSOs.  The
horizontal dashed line indicates the absolute $I$-band magnitude of an L*
galaxy.  The 2MASS QSO host magnitudes are consistent with the 
locus formed by the comparison sample, and both groups have a mean value
of $\sim$1.5 L*.

On the otherhand, the 2MASS nuclear luminosities are generally fainter, ranging 
over seven magnitudes.  Several of the brightest objects are as bright as the
PG QSOs.  The remaining majority do not satisfy the traditional
criterion for QSO classification as indicated by
the vertical dashed line in Figure~\ref{fig_agn_vs_gal}.  Recall from 
Figure~\ref{fig_pg} that 2MASS 
QSOs are all as bright as PG QSOs in the $K$-band.  If the 2MASS and PG QSOs 
also share similar intrinsic nuclear $I$-band
luminosities, then between zero and seven ($mean =2.7$) magnitudes
of extinction in the $I$-band are required to shift their luminosities to 
the average of the PG sample, $\langle M_I \rangle=-23.7$.  

\begin{figure*}[t]
\epsscale{.45}
\vskip 12pt
\hskip 110pt
\plotone{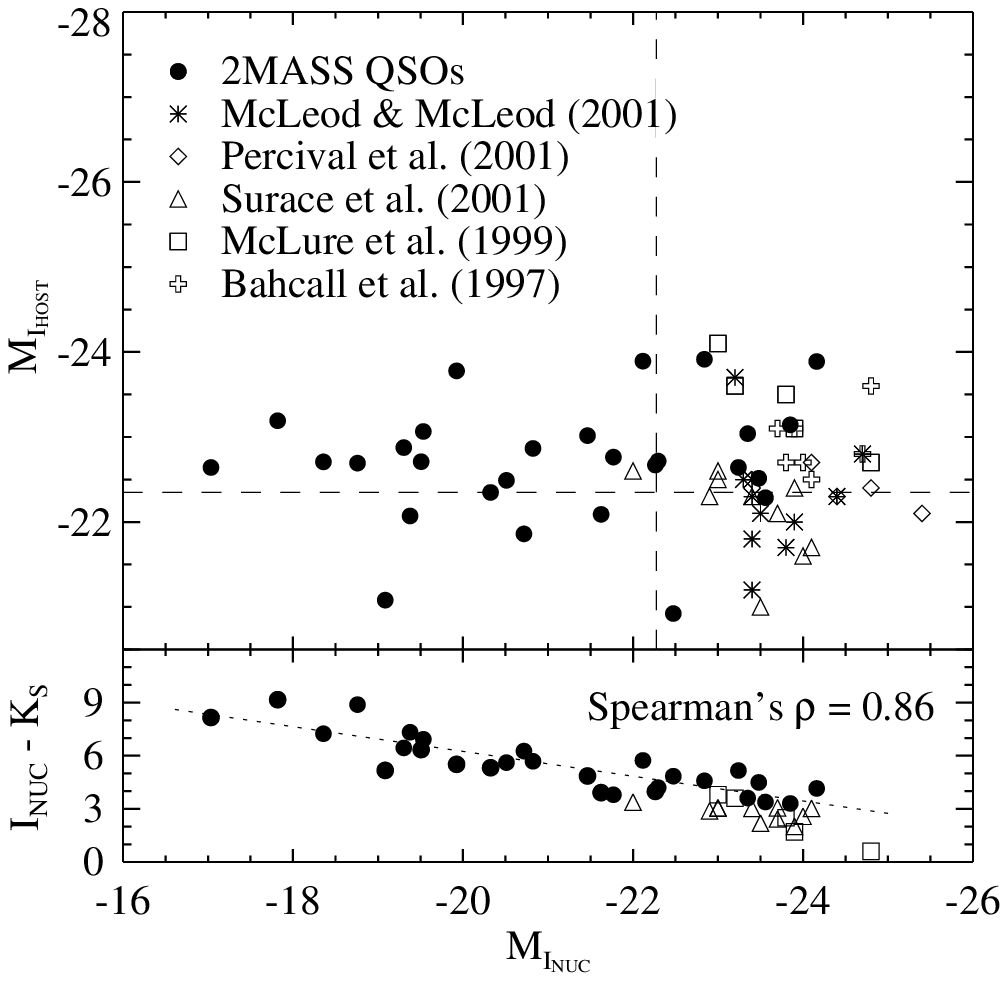}
\vskip 6pt
\caption{(top) Host galaxy vs nuclear luminosities. 
$M_I$ for an L* galaxy ($-22.3$) and the historical $M_B<-22.1+5 \log h$
QSO criterion are indicated by dashed lines.
(bottom) Nuclear luminosity versus $I-K_S$
color ($K_S$ includes nuclear and host light). 
The dotted line is a linear fit to the 2MASS QSO 
data points only.}
\end{figure*}

The bottom panel of Figure~\ref{fig_agn_vs_gal} plots the 2MASS QSO nuclear 
luminosities versus their $I_{NUC}-K_S$ colors.  A Spearman's rank correlation 
test ($\rho=0.86$) confirms the correlation suggested by the linear fit 
(dotted line).  Some of the scatter is due to the fact that the $K_S$ 
values plotted are total magnitudes including both nuclear light and light
from the host galaxy.  The fact that dimmer 2MASS QSO nuclei are also
redder is consistent with dust obscuration causing the large range in 
nuclear luminosities.  There appears to be no relationship between
morphology and $I_{NUC}-K_S$ or $M_{I_{NUC}}$, suggesting that any obscuring
material is local to the nucleus and not dependent on the host galaxy.
Where suitable color information was available, PG QSO data points have
been included on the plot.

Various studies in recent years have suggested that more luminous nuclei
are preferentially found in elliptical hosts. Among the samples considered 
in this paper, this claim is supported
by the results of \citet{mclu99} and contradicted by those of \citet{per01}.
We find no correlation between host galaxy morphology and the luminosity of
the nucleus for the 38 radio-quiet PG QSOs, nor for the 2MASS QSOs 
($K_S$-band).

Galaxy interaction, a possible fueling mechanism for AGN, is an important parameter 
to consider when comparing QSO samples.  However, visual inspection of a 
host galaxy's dynamical state is subject to a number of problems.  Increased
star formation triggered by minor and major mergers is more highly contrasted in bluer filters,
complicating comparisons of studies made at different wavelengths.  Additionally,
the sensitivity and resolution of the instrument used for observations affects
whether or not faint tidal tails are detected or multiple nuclei are 
distinguished.  Of course, the subjectivity of assessing degrees of interaction
is also sensitive to both the investigator and the investigator's working
definition of ``interaction''.  

Tables~\ref{tbl_properties} and~\ref{tbl_rqq} list both the 2MASS QSOs and 
radio-quiet PG QSOs as exhibiting or not exhibiting evidence of interaction.
This sometimes uncertain distinction was made by a single individual to 
remove some degree of subjectivity.  Criteria for interaction include 
multiple nuclei, obviously merging galaxies, remnant debris such as tidal tails,
and clear asymmetry.  Ten of the 2MASS QSOs (34\%) were classified as past
or present interacting systems, as were ten of the radio-quiet PG QSOs (29\%).
Considering the inhomogeneity of the PG imaging, rough agreement between the 
two samples seems reasonable.

A more generic indicator of the host galaxy environment is the number of companion
galaxies within a fixed physical distance.  In the absence of known redshifts, all 
apparent companions (extended objects with small projected separations) can be 
considered for comparative purposes.  The methodology described in 
\S \ref{sec_env} was chosen in order to duplicate the \citet{bah97} study of
apparent companions around a sample of radio-quiet PG QSOs.  Only those 
objects with projected separations of 25 h$^{-1}$ kpc or less are presented
by those authors.  However, this coincides with the maximum distance at which 
apparent companion information is complete for both the 2MASS QSOs and the QDOT 
ULIRGs drawn from the \emph{HST} archive.  Thus, a fair comparison is possible 
between these three samples.

\begin{figure*}[t]
\epsscale{.45}
\vskip 12pt
\hskip 5pt
\plotone{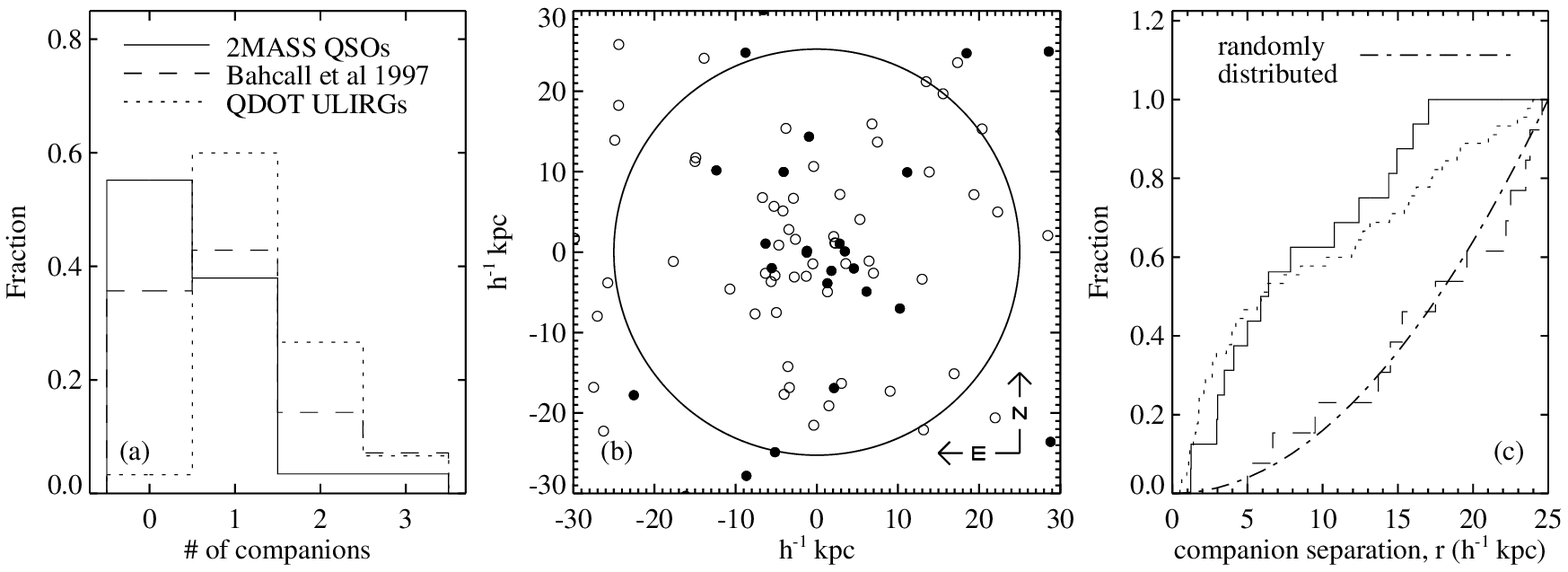}
\vskip 6pt
\caption{Comparison of 2MASS QSO, radio-quiet PG QSO and ULIRG apparent companions with 
$M_I \leq -18.3$ and $0\leq r \leq25$ $h^{-1}$ kpc.
(a) Apparent companion frequency (fraction of sample with zero, 
one, two or three projected companions); one ULIRG with five is not
shown here. (b) Positions of apparent companions relative to QSO
or peak brightness.  Filled circles correspond to 2MASS QSOs 
and unfilled circles represent ULIRGs.  The large circle
indicates the completeness limit, $r = 25$ $h^{-1}$ kpc. (c) 
Cumulative distributions of projected separations for all three
samples as well as the expectation for galaxies randomly distributed in 
the plane of the sky.}
\end{figure*}

\begin{table}[b]
\begin{center}
\vskip 12pt
{\sc Table \ref{tbl_ks}\\
\smallskip\hbox to\hsize{\hfil{Kolmogorov-Smirnov statistics\tablenotemark{*}}\hfil}}
\small
\setlength{\tabcolsep}{6pt}
\begin{tabular}{rccc}
\tskip \tableline \tableline \tskip
\multicolumn{4}{c}{Apparent Companion Frequencies} \\
\tskip \tableline \tskip
                       & B97\tablenotemark{a}   & ULIRG\tablenotemark{b}    &                      \\
2MASS\tablenotemark{c} & $82.5$                 & $0.03$                    &                      \\
B97                    &          \nodata       & $0.32$                    &                      \\
\tskip \tableline \tskip
\multicolumn{4}{c}{Apparent Companion Separations} \\
\tskip \tableline \tskip
	               & B97                    & ULIRG                     & RND\tablenotemark{d} \\
2MASS                  & $2.59$                 & $50.1$                    & $0.01$               \\
B97                    &         \nodata        & $1.60$                    & $79.0$               \\
ULIRG                  &         \nodata        &         \nodata           & $10^{-7}$            \\
\tskip\tableline
\end{tabular}
\vspace*{-12pt}
\tablenotetext{*}{The KS test is a two-sided non-parametric statistic which returns the normally-distributed probability that two samples were drawn from the same parent population.  Values in this table are presented as percentages.}
\tablenotetext{a}{\citet{bah97}}
\tablenotetext{b}{QDOT ULIRGs (this paper)}
\tablenotetext{c}{2MASS QSOs (this paper)}
\tablenotetext{d}{random distribution}
\end{center}
\end{table}

Figure~\ref{fig_companions}a shows the apparent companion frequencies for each
sample.  The number of 2MASS QSOs, radio-quiet PG QSOs and
ULIRGs with no apparent companions is $16/29$ ($55\%$), $5/14$ ($36\%$), and 
$1/30$ ($3\%$), respectively.  Not shown on this plot is one ULIRG
with five such objects.  Table~\ref{tbl_ks} lists results of KS tests 
applied to the frequency distributions as well as the projected 
separations. These values indicate the probability (in percentage form) that
two samples are drawn from the same parent distribution.  
The number of apparent companions around the 2MASS QSOs is not 
statistically different than the number around the radio-quiet PG QSOs.  
However, the 2MASS QSO and ULIRG distributions differ at the 99.97\% 
confidence level.

Figure~\ref{fig_companions}b shows the positions of apparent companions relative 
to the central object
in both the 2MASS QSO and ULIRG samples.  The former are indicated by
filled circles.  Given the similar samples sizes (29 and 30), the far 
greater number of apparent companions in the ULIRG images is clear.

The projected separations are binned and plotted as
cumulative histograms for all three samples in Figure~\ref{fig_companions}c.
Additionally, the profile expected for objects that are randomly 
distributed in the plane of the sky is shown.  Both the 2MASS QSOs and the ULIRGs 
show an excess over random at close separations.  The cumulative distribution
rises sooner for the latter, reflecting smaller projected separations of the
closest extended objects. The probability of being randomly distributed in the 
plane of the sky is $10^{-9}$ and $10^{-3}$ for the ULIRG and the 2MASS QSO 
apparent companions, respectively.  The apparent companions identified by 
\citet{bah97}, for their subset of radio-quiet PG QSOs, cannot be distinguished
from the random expectation (with respect to radial distribution only, not total 
number).

Interpretation of the comparative apparent companion results is not clear-cut.  The
radial distribution implies a similarity between the 2MASS QSO and the ULIRG
environments.  However, if 2MASS QSOs represent evolved ULIRGs, one might expect
to find more closely separated companions around the former.  The opposite is in fact seen.
The random radial distribution of apparent companions around the PG QSOs may or may not 
indicate that previously close companions have already been swallowed by the 
host galaxies.  What is clear is that the ULIRGs studied have many more apparent 
companions at all separations (within 25 $h^{-1}$ kpc) than either the 2MASS QSOs
or the radio-quiet PG QSOs, and unlike the other two samples, the ULIRGs
nearly universally have at least one apparent companion.

The immediate environments of the ULIRGs and 2MASS QSOs are 
significantly different. \citet{far01} were able to classify morphologies 
for only two of 23 ULIRGs (both ellipticals).  All but one of the objects 
in their sample are obviously interacting systems.  Similarly, \citet{sur98} found
that eight of the nine objects in their sample exhibit evidence that they are
advanced merger systems.  These results are consistent with those of \citet{vei02}
who found that $\sim73\%$ of their ULIRGs could be adequately fit with a 
de Vaucouleurs profile and all but one of the 118 ULIRGs show signs of strong 
interation.   The remains of rapid
star formation and universal tidal debris that one might expect to see if
the red QSOs found by 2MASS represent an intermediate stage of evolution 
between ULIRGs and PG QSOs do not appear to be present.  Clearly, many of
the 2MASS QSOs have undergone or are undergoing interaction events.  
However, many others appear to be isolated, undisturbed galaxies.

There is growing evidence that 2MASS QSOs may simply be explained by some form of the 
proposed unified model for AGN.  In that scenario, these objects are normal QSOs that
are heavily obscured along our line of sight.  \citet{smi02} found the sample as a whole
to be more highly polarized than other QSO samples.  We have shown by properly accounting
for starlight dilution, that the polarization of the nuclear region is significantly 
underestimated by broad-band polarimetry in many cases. Furthermore, the 
magnitude of the corrections is correlated to AGN type, with the nucleus to galaxy flux ratio 
generally smaller for narrow-line objects.   All of this combined
with the lack of narrow polarized emission lines in many of the highly polarized objects
\citep{smi00,smi03,sch02}
is consistent with dust obscuration near the nucleus.  If obscuring dust were generally
located further away, as in a galactic disk, one 
might expect $I_{NUC}-K_S$ to correlate with morphology.  We see no evidence of this.

The fact that the 2MASS QSOs have similar absolute $K_S$-band magnitudes
 and that their host galaxies
do not distinguish themselves significantly from those of PG QSOs in
the $I$-band implies that the red colors observed are a property of the
nucleus alone.  The 2MASS AGN are dominated by
type 1 objects, and edge-on host galaxies seem to be under-represented.
Therefore, even the 2MASS AGN sample is probably missing extremely obscured
objects like those identified by \emph{IRAS}.  In these putative highly obscured
objects, the AGN may not contribute significantly in the $J$-band.  In that
case, the $J$-band flux will reach a minimum consistent with the host galaxy
flux, and these objects will not have easily distinguishable $J-K_S$ colors.
The AGN might become slightly visible in $K_S$, but may still not distinguish
itself.

Chandra observations show that many 2MASS AGN are faint (probably absorbed)
in x-rays \citep{wil02}, hindering the identification of highly
obscured AGN in the nearby universe.  The presence of
extinguishing dust very near the nucleus, as the evidence suggests for the
2MASS QSOs, should make these objects bright in the mid-to-far infrared
compared with their optical/near-IR emission.  While this question will be
examined for a large number of 2MASS AGN with SIRTF, identification of many
highly obscured QSOs in the local universe may have to wait for future, deep
all sky surveys at wavelengths longer than two microns.

\section{Conclusions}\label{sec_conclusions}

We have observed and analyzed a sample of 29 red QSOs selected from the 
2MASS near-infrared survey.  \emph{HST}/WFPC2 $I$-band images were used to 
model the PSF and subtract nuclear light to study the host galaxies.  
We find that 2MASS QSOs live in a variety of galaxy types with luminosities
ranging between 0.3 and 4.2 L*.  Of the 26 
objects with obvious hosts, 9 were found in spiral galaxies and 4 in 
ellipticals.  The remaining sixteen could not be classified with
confidence.  It was argued that at least half of these may be spirals 
based on the distribution of measured eccentricities.
Roughly a third of the sample exhibit evidence of interaction, predominantly 
those with ambiguous morphologies.

The immediate environments of the 2MASS QSOs were inspected by identifying 
apparent companions with projected separations less than 25 $h^{-1}$ kpc.  Less 
than half the sample has one or more apparent companions, with one having two and
another three.  When compared to a sample of radio-quiet PG QSOs, no statistically 
significant difference was found.  This is in contrast to the same analysis
performed on a sample of ULIRGs for which all but one object had at least 
one apparent companion.  The projected radial distributions suggest that companion
galaxies lie closer, on average, to the 2MASS QSOs and ULIRGs than the PG QSOs,
though data is available for fewer objects in the latter sample.

The 2MASS QSOs were already known to be a relatively highly polarized 
sample.  The amount of unpolarized starlight from host galaxies 
was measured and used to estimate polarization corrections.  The resulting 
increase is substantial and depends on AGN type, with type 2s generally having the 
largest corrections.

Absolute $I$-band magnitudes of the AGN span the range
$-17.0$ to $-24.2$.   A strong correlation between $M_{I_{NUC}}$ and $I_{NUC}-K_S$ 
color was noted, with redder objects appearing fainter.  Neither nuclear luminosity 
nor $I_{NUC}-K_S$ was found to correlate with morphology.  

No indication was seen that the red 2MASS QSOs
represent an intermediate evolutionary step between ULIRGs and classical QSOs.  
In light of the similarities between host galaxies and environments, the results of this study
suggest that the relatively faint 2MASS QSO $I$-band nuclear luminosities are caused
by obscuration local to the nucleus and that the 2MASS QSOs differ from radio-quiet
PG QSOs only by their orientation with respect to the observer.  Thus, the
red QSOs being identified by 2MASS are likely the missing QSO population predicted
to exist by x-ray and radio surveys.
However, the range of properties observed leaves open the 
possibility that more than one phenomenon is responsible for the nature of 
the large number of QSOs discovered by 2MASS.

\acknowledgments
	
We are grateful to Marcia Rieke for providing galaxy SED templates that extend 
into the near-infrared.  Thanks to Dennis Zaritsky, Chris Impey, George Rieke,
and the referee for insightful comments and suggestions.  This publication
makes use of data products from the Two Micron All Sky Survey, which is a joint
project of the University of Massachusetts and the Infrared Processing and 
Analysis Center/California Institute of Technology, funded by NASA and the NSF.
Support for this work was provided by NASA through
grant number GO-9057 from the Space Telescope Science Institute, which is
operated by Association of Universities for Research in Astronomy,
Incorporated, under NASA contract NAS5-26555.

\appendix

\section{Comments on Individual Objects}\label{app_notes}

\emph{2M000810+1354}$-$Seemingly isolated spiral galaxy with only $\sim$1\% of total 
                       observed light coming from the nucleus.

\emph{2M005055+2933}$-$Either a modestly inclined barred spiral galaxy with visible 
                       dust obscuration or an atypical edge-on warped disk.
                       Observed light is dominated by host galaxy, and the corrected
	               polarization exceeds 13\%.

\emph{2M010607+2603}$-$Highly polarized object with very little relative light contribution 
                       from the morphologically ambiguous host galaxy ($z\sim0.4$).  The
                       surface brightness profile is very poorly determined; however, the
                       host must either be a highly inclined spiral or an interacting 
                       system.  The polarization angle is closely aligned in either case.

\emph{2M015721+1712}$-$Asymmetrical host galaxy responsible for roughly half of the total light.
                       An extended feature in southeast corner is present both before and after 
                       PSF-subtraction.  A possible faint companion or tidal debris is seen 
                       within 6 projected $h^{-1}$ kpc below the object.

\emph{2M022150+1327}$-$Although not visible in Figure~\ref{fig_hosts1}, an apparent tidal
                       arm (exceeding 20 $h^{-1}$ kpc in length) extends from the bottom 
                       of the host image and arcs to the north.  Also, there is
                       marginal visual evidence that the host galaxy is in fact two 
                       merging galaxies of equal size.

\emph{2M023430+2438}$-$Residual image and surface brightness profile yield essentially no
                       information about the $\sim$2 L* host galaxy at $z\sim0.3$.

\emph{2M032458+1748}$-$Smoothly distributed emission from host galaxy with majority of total
                       observed light originating from the nucleus.

\emph{2M034857+1255}$-$Shares the distinction (with 2M130700+2338) of having reddest near-IR
                       color ($J-K_S=3.3$) and clearly comprises two merging galaxies.

\emph{2M092049+1903}$-$Apparently isolated spiral galaxy in relatively highly populated field.
                       Only one apparent companion meets the adopted magnitude cutoff, although 
                       six additional extended objects have projected separations of 25 $h^{-1}$
                       kpc or less.

\emph{2M125807+2329}$-$Very faint elongated host galaxy centered roughly on the QSO nucleus.

\emph{2M130700+2338}$-$Elliptical galaxy with a very red near-IR color ($J-K_S=3.3$ for AGN+host).

\emph{2M145331+1353}$-$Clear spiral structure with large observed light contribution from 
                       host galaxy.

\emph{2M151621+2259}$-$Very faint spiral arms emerging from a possible bar.

\emph{2M152151+2251}$-$Disturbed host showing asymmetry and a luminous knot of unknown nature.

\emph{2M154307+1937}$-$Although fairly well described by a $r^{1/4}$ profile, residual flux
                       exhibits obvious asymmetry.  There is a possible alignment of the 
                       polarization angle and the fitted isophotes, however the degree of 
                       polarization is very low.

\emph{2M163700+2221}$-$Host galaxy possesses a disk with multiple luminous knots within 5
                       $h^{-1}$ kpc of nucleus.  Not evident in Figure~\ref{fig_hosts1} is 
                       possible tidal debris extending from the host in the northwest direction.
                       The corrected polarization is $\sim$7\% and the polarization angle is
                       closely aligned with the disk.

\emph{2M163736+2543}$-$An apparently merging system with the majority of observed emission
                       from the galaxies.  Close inspection of the inner region reveals 
                       multiple nuclei.

\emph{2M165939+1834}$-$Inner region clearly resolved into two distinct components.  Slightly
                       more than half of the observed luminosity comes from the nucleus
                       which has a corrected polarization of $\sim$11\%.  Interestingly,
                       the polarization angle is rotated almost exactly $90\degr$ from the
                       axis through the two components, perhaps suggesting scattering of
                       nuclear light by a close companion.

\emph{2M170003+2118}$-$No residual flux is visible to reveal the nature of the $\sim$L*
                       galaxy hosting the highest redshift QSO in the sample ($z\sim0.6$).
                       The corrected polarization is $\sim$11\%.

\emph{2M171442+2602}$-$Grand design spiral with three apparent companions.  The anomalous
                       light trail is identically present in both 400 second integrations.

\emph{2M171559+2807}$-$Remarkable asymmetry present in PSF-subtracted image.  The high 
                       polarization and orientation of the polarization angle are consistent
                       with the extended feature being an ionization cone.

\emph{2M222202+1952}$-$Point-like object used for construction of the PSF (this explains the
                       odd appearance of the PSF-subtracted image in Figure~\ref{fig_hosts1}).

\emph{2M222221+1959}$-$Faint, smoothly distributed residual flux with disturbed surface
                       brightness profile.

\emph{2M222554+1958}$-$Inner portion of spiral arms seen in Figure~\ref{fig_hosts1}.  Not
                       visible is faint extension of these arms.  Only $\sim$1\% of the 
                       total observed light is from the nucleus.

\emph{2M225902+1246}$-$Excellent agreement with $r^{1/4}$ profile and smooth visual 
                       appearance.  Recent observations have established that this
                       object's $J-K_S$ color barely falls short of the 2MASS AGN criterion,
                       though it was included in the full analysis.

\emph{2M230304+1624}$-$Clearly an interacting system.  In addition to the extremely close
                       projected separation of the two galaxies, a faint luminous tail
                       extends towards the northeast corner of the image in 
                       Figure~\ref{fig_hosts1}.

\emph{2M230442+2706}$-$Surface brightness profile described well by $r^{1/4}$ law despite
                       interference beyond 2.5 $h^{-1}$ kpc by apparent companion with $\sim$5 
                       $h^{-1}$ kpc projected separation.

\emph{2M232745+1624}$-$Faint apparent spiral arms.  Inclination may have been overestimated
                       if host galaxy is in fact a barred spiral.

\emph{2M234449+1221}$-$Very little visible residual flux around PSF-subtracted nucleus.
                       Well behaved surface brightness profile is misleading as it probes
                       different objects at different radii.  The system appears to be 
                       interacting.  A common envelope seems to include the QSO and
                       the extremely close companion object.  The polarization angle is
                       rotated nearly $90\degr$ from the axis connecting the two bright spots.

\end{document}